%% file: main.tex
\IfFileExists{venue_preamble.tex}{\def\VenueMode{1}
  \input{venue_preamble}
}{\documentclass[11pt]{article}
  \usepackage[margin=1in]{geometry}
}

\input{preamble_common}
\input{macros}
\input{figures/style}
\input{metadata}

\title{\PaperTitle}
\author{\PaperAuthors}
\date{\PaperDate}
\hypersetup{pdftitle={\PaperTitle},pdfauthor={\PaperPdfAuthors}}

\providecommand{\PaperMakeTitle}{\maketitle}
\providecommand{\PaperPrintBibliography}{\ifPaperPrintReferences
    \bibliographystyle{alpha}
    \bibliography{references}
  \fi}
\providecommand{\PaperEndMatter}{}

\begin{document}

\PaperMakeTitle

\begin{abstract}
\input{sections/00_abstract}
\end{abstract}

\clearpage

\input{sections/01_introduction}
\input{sections/02_overview}
\input{sections/03_preliminaries}
\input{sections/04_rectangles}
\input{sections/05_pair_shuffles}
\input{sections/06_level_words}
\input{sections/07_consequences}

\PaperPrintBibliography

\input{sections/11_ai_methodologies}

\PaperEndMatter

\end{document}

%% file: preamble_common.tex
\usepackage[T1]{fontenc}
\usepackage{microtype}
\usepackage{amsmath}
\usepackage{amssymb}
\usepackage{amsthm}
\usepackage{aliascnt}
\usepackage{mathtools}
\usepackage{graphicx}
\usepackage{booktabs}
\usepackage{array}
\usepackage{xcolor}
\usepackage{tikz}
\usetikzlibrary{arrows.meta,backgrounds,calc,decorations.markings,decorations.pathreplacing}
\definecolor{DPBlue}{HTML}{0072B2}
\definecolor{DPOrange}{HTML}{D55E00}
\definecolor{DPPurple}{HTML}{7057A3}
\tikzset{
  dp one/.style={draw=DPBlue,line width=1.1pt},
  dp two/.style={draw=DPOrange,line width=1.1pt},
  dp edge/.style={draw=DPPurple,line width=1.6pt,-{Stealth[length=4pt]}},
  dp arrow/.style={postaction={decorate},decoration={markings,mark=at position 0.58 with {\arrow{Stealth[length=4pt]}}}},
  dp label/.style={inner sep=1.5pt,font=\normalsize}
}
\usepackage{algorithm}
\usepackage{algpseudocode}
\algrenewcommand\algorithmicindent{1.3em}
\algrenewcommand\algorithmiccomment[1]{\hfill$\triangleright$ #1}
\usepackage{hyperref}
\usepackage[nameinlink,capitalize,noabbrev]{cleveref}

\ifdefined\VenueMode
\else
  \hypersetup{
    colorlinks=true,
    citecolor=blue,
    linkcolor=blue,
    urlcolor=blue
  }
\fi

\graphicspath{{figures/}}
\allowdisplaybreaks

%% file: macros.tex
\newcommand{\OPT}{\mathrm{OPT}}
\newcommand{\Funnel}{\mathrm{Funnel}}
\newcommand{\add}{\mathrm{add}}

\newcommand{\err}{\mathrm{err}}

\newcommand{\desc}{\mathrm{desc}}

\newcommand{\Pts}{\mathrm{pts}}
\newcommand{\IRB}{\mathrm{IRB}}
\newcommand{\IB}{\mathrm{IB}}
\newcommand{\Alt}{\mathrm{Alt}}
\newcommand{\rect}{\mathrm{box}}

\newcommand{\occ}{\mathrm{occ}}
\newcommand{\tops}{\mathrm{top}}
\DeclarePairedDelimiter{\abs}{\lvert}{\rvert}
\DeclarePairedDelimiter{\ceil}{\lceil}{\rceil}

\theoremstyle{plain}
\newtheorem{theorem}{Theorem}[section]
\newaliascnt{lemma}{theorem}
\newtheorem{lemma}[lemma]{Lemma}
\aliascntresetthe{lemma}
\newaliascnt{proposition}{theorem}
\newtheorem{proposition}[proposition]{Proposition}
\aliascntresetthe{proposition}
\newaliascnt{corollary}{theorem}
\newtheorem{corollary}[corollary]{Corollary}
\aliascntresetthe{corollary}

\theoremstyle{definition}
\newaliascnt{definition}{theorem}
\newtheorem{definition}[definition]{Definition}
\aliascntresetthe{definition}

\theoremstyle{remark}
\newaliascnt{remark}{theorem}
\newtheorem{remark}[remark]{Remark}
\aliascntresetthe{remark}

%% file: figures/style.tex
\tikzset{
  figtxt/.style={font=\scriptsize,inner sep=1pt},
  sdot/.style={circle,fill=black!70,inner sep=0pt,minimum size=4.2pt},
  gdot/.style={circle,draw=DPOrange,fill=white,line width=0.9pt,inner sep=0pt,minimum size=4.6pt},
  tdot/.style={circle,fill=DPOrange,inner sep=0pt,minimum size=5pt},
}
\tikzset{
  rst/.style={draw=DPBlue,line width=0.9pt,fill=DPBlue,fill opacity=0.10,draw opacity=1},
  rplus/.style={draw=DPOrange,line width=0.9pt,dashed,fill=DPOrange,fill opacity=0.12,draw opacity=1},
  rin/.style={draw=DPPurple,line width=1.1pt,densely dotted,fill=DPPurple,fill opacity=0.10,draw opacity=1},
  cdot/.style={circle,draw=black,fill=white,line width=0.8pt,inner sep=0pt,minimum size=5.6pt},
  varrow/.style={draw=black!60,line width=0.5pt,-{Stealth[length=3.2pt]},shorten >=2.4pt,shorten <=2.4pt},
}
\newcommand{\legendbox}[1]{\tikz[baseline=-0.55ex]\draw[#1] (0,-0.13) rectangle (0.5,0.17);}

%% file: metadata.tex
\newcommand{\PaperTitle}{Wilber's Funnel Bound Is Not Tight}
\newcommand{\AuthorCao}{Yang Cao}
\newcommand{\AuthorQi}{Haoran Qi}
\newcommand{\AuthorWang}{Hanzhi Wang}
\newcommand{\PaperAuthors}{\AuthorCao\thanks{\texttt{yangcao@psu.edu}. Pennsylvania State University.} \and \AuthorQi\thanks{Jilin University.} \and \AuthorWang\thanks{Tsinghua University.}}
\newcommand{\PaperPdfAuthors}{\AuthorCao, \AuthorQi, \AuthorWang}
\newcommand{\PaperDate}{}

\newif\ifPaperPrintReferences
\PaperPrintReferencestrue

%% file: sections/00_abstract.tex
Wilber [SIAM J.\ Comput.\ 1989] gave two lower bounds on the cost of binary search trees and conjectured that the second one, the funnel bound, is tight: on every sequence of searches, the optimal cost is within a constant factor of the funnel bound. We refute this conjecture. For infinitely many $m$, we construct a sequence of $m$ searches on $m$ distinct keys whose funnel bound is $O(m)$, while every binary search tree, even an offline one, has cost $\Omega(m\log\log m)$ on it. Among sequences of $m$ searches on $m$ distinct keys, this gap is the largest possible. On the same sequences, the independent rectangle bound of Demaine et al.\ [SODA 2009] exceeds the funnel bound by a factor $\Omega(\log\log m)$. This refutes a conjecture of Iacono that the two bounds are equivalent. Moreover, the two signed Greedy bounds of Demaine et al., one for each orientation, differ by $\Omega(m\log\log m)$. This refutes a conjecture of Lecomte and Weinstein [ESA 2020] that they differ by only $O(m)$.

%% file: sections/01_introduction.tex
\section{Introduction}
\label{sec:intro}

A binary search tree (BST) on a set of keys serves a sequence $X\coloneqq x_1\cdots x_m$ of searches. Each search starts at the root and must reach the node with the searched key; along the way, the tree may be rearranged by rotations. In the model of Demaine, Harmon, Iacono, Kane and P\u{a}tra\c{s}cu \cite{dhi+09}, which is equivalent to the usual models up to constant factors, the $t$-th search touches a connected subtree that contains the root and $x_t$, rearranges the touched nodes into any BST on them, and costs the number of touched nodes. Let $\OPT(X)$ be the least total cost of serving $X$ by a BST that knows $X$ in advance and starts from the best initial tree. Sleator and Tarjan conjectured that splay trees serve every sequence within a constant factor of $\OPT(X)$ \cite{st85}. This dynamic optimality conjecture is open. The best competitive ratio known for an online BST on $n$ keys is $O(\log\log n)$, achieved by Tango trees \cite{dhip07} and multi-splay trees \cite{wds06}. Recent preprints prove competitive ratios of $O(\log\log n\cdot(\log\log\log n)^2)$ for splay trees \cite{chh+26} and $2^{O(\sqrt{\log\log n})}$ for Greedy \cite{gpsw26}, another candidate dynamically optimal BST introduced independently by Lucas \cite{luc88} and Munro \cite{mun00} and cast in geometric form by Demaine et al.\ \cite[Section~3]{dhi+09}.

All these results compare an algorithm with a lower bound on $\OPT$, since $\OPT$ itself is not known to be computable efficiently. A lower bound that is within a constant factor of $\OPT$ on every sequence would reduce dynamic optimality to showing that an online BST matches it. Wilber \cite{wil89} gave the first two such bounds. The first, the \emph{alternation bound}, fixes a static reference tree and counts, at every node, how often consecutive searches below that node switch between its left and right subtrees. Tango trees are $O(\log\log n)$-competitive because their cost is within this factor of the alternation bound \cite{dhip07}, and Jiang, Lecomte, Weinstein and Yingchareonthawornchai observe that every $o(\log n)$-competitive BST known to them is analyzed against this bound \cite{jlwy24}.

The second, the \emph{funnel bound}, needs no reference tree. Draw each search $x_t$ as the point $(x_t,t)$ in the plane, with keys ordered from left to right and time increasing upwards. The \emph{funnel} of a search $p$ consists of the earlier searches $q$ with a different key such that the closed axis-parallel rectangle with corners $p$ and $q$ contains no other search; each such $q$ lies either to the left or to the right of $p$, forming the \emph{left funnel} and \emph{right funnel} of $p$. The funnel bound charges $p$ the number of maximal runs of searches on the same side when its funnel is listed in order of time, and $\Funnel(X)$ is the sum of these charges over all searches. (\Cref{sec:prelim-funnel} gives the precise definition, due to Lecomte and Weinstein \cite{lw20}.) Both bounds are lower bounds on $\OPT$ up to a constant factor and an additive $O(m)$ \cite{wil89}. The funnel bound was used to prove key-independent optimality \cite{iac05}, and Levy and Tarjan related it to the amortized analysis of splay trees \cite{lt19}.

Which of these bounds is close to $\OPT$? Wilber conjectured, in the conference version (FOCS 1986) of \cite{wil89}, that the second bound is at least as strong as the first, and that it is tight; in the formulation of Lecomte and Weinstein, the second conjecture states that $\Funnel(X)=\Theta(\OPT(X))$ for every access sequence $X$ \cite[Section~1]{lw20}. Two later lower bounds are at least as large as both of Wilber's bounds, up to constant factors. Derryberry, Sleator and Wang gave one \cite{dsw05}, \cite[Section~3.5.1]{der09}, and Demaine et al.\ gave the \emph{independent rectangle bound} $\IRB(X)$ \cite{dhi+09}, the largest size of a set of empty rectangles (closed rectangles spanned by two searches with different keys that contain no other search) that are pairwise independent, in the sense that no rectangle contains a corner of another in its interior. They proved $\OPT(X)\ge\IRB(X)/2+m$ \cite[Theorem~4.1]{dhi+09}, showed that $\IRB(X)$ is at least as large as both of Wilber's bounds up to constant factors \cite[Section~4.1]{dhi+09}, and gave a greedy algorithm, \emph{signed Greedy}, that computes it up to a constant factor (we describe it below). They describe the expectation that Wilber's second bound is stronger than the first, and possibly within a constant factor of $\OPT$, as folk wisdom going back to Wilber's paper, and they suggest that the best independent rectangle bound may also be within a constant factor of $\OPT$ \cite[Section~1]{dhi+09}. Since the independent rectangle bound is at least as large as the funnel bound up to constant factors, the question is whether it can exceed the funnel bound by more than a constant factor. Iacono conjectured that it cannot, that is, that the two bounds are equivalent \cite[Section~3]{iac13}.

Recent work confirmed the first of Wilber's two conjectures. Lecomte and Weinstein proved that the alternation bound of every reference tree is $O(\Funnel(X)+m)$ \cite[Theorem~1]{lw20}, and that it can be smaller by a factor $\Theta(\log\log n)$ \cite[Theorem~2]{lw20}; Jiang et al.\ amplified such separations by composing sequences \cite{jlwy24}. The alternation bound is not tight: Chalermsook, Chuzhoy and Saranurak \cite{ccs23} and Lecomte and Weinstein \cite{lw20} constructed sequences on which it is smaller than $\OPT$ by a factor $\Omega(\log\log n/\log\log\log n)$ and $\Theta(\log\log n)$, respectively, for every choice of the reference tree. Whether the funnel bound is tight remained open, and Wilber's second conjecture has been repeated since \cite{lw20,jlwy24}. Lecomte and Weinstein describe the equivalence of the funnel bound and the independent rectangle bound as strongly believed, and note that it would help to analyze candidate optimal BSTs such as splay trees and Greedy \cite[Section~5]{lw20}. Guo, Pettie, Skora and Wan recently noted that it is still unknown whether the lower bounds of Demaine et al.\ can exceed the funnel bound, or the bound of Derryberry, Sleator and Wang, by more than a constant factor \cite[Section~1.1]{gpsw26}. So the conjectures in the literature point the same way: the funnel bound is within a constant factor of $\OPT$ \cite{wil89,lw20,jlwy24}, and it agrees with the independent rectangle bound up to a constant factor \cite{iac13,lw20}. If the funnel bound were within a constant factor of $\OPT$, the dynamic optimality of an online BST such as splay trees or Greedy would follow from a bound $O(\Funnel(X)+m)$ on its cost. Unlike a bound by the alternation bound, such a bound would involve no reference tree.

We refute Wilber's conjecture that the funnel bound is tight, and with it the equivalence of the funnel bound and the independent rectangle bound (\cref{cor:irb-informal}).

\begin{theorem}[Informal version of \cref{thm:main-formal}]
\label{thm:main-informal}
For every integer $k\ge1$ there is a permutation $\pi_k$ of $m$ keys, where $m$ tends to infinity with $k$, such that
\begin{align*}
\Funnel(\pi_k)=O(m),\qquad\OPT(\pi_k)=\Omega(m\log\log m).
\end{align*}
\end{theorem}

So the optimum exceeds the funnel bound by a factor $\Omega(\log\log m)$. With explicit constants, \cref{thm:main-formal} gives $\Funnel(\pi_k)\le7m$, $\OPT(\pi_k)\ge(1+k/2)m$ and $\log_2\log_2m\le2k+2$. Since $\Funnel(\pi_k)=O(m)$ while $\OPT(\pi_k)/m\to\infty$, the theorem holds for every variant of the funnel bound that differs from ours by a constant factor and an additive $O(m)$, which covers the variants in the literature, including Wilber's original definition \cite[footnote~6]{lw20}. It also holds for the funnel bound of every image of $\pi_k$ under the symmetries of the square, acting on the $m\times m$ grid of keys and times (\cref{rem:symmetries}). For permutations of $m$ keys, the gap $\Theta(\log\log m)$ is the largest possible:

\begin{proposition}[Informal version of \cref{prop:tango-formal}]
\label{prop:tango-informal}
For every permutation $X$ of $m\ge4$ keys, $\OPT(X)\le O(\log\log m)\cdot\Funnel(X)$.
\end{proposition}

This follows from the analysis of Tango trees and the comparison of Wilber's two bounds by Lecomte and Weinstein \cite[Theorem~1]{lw20}; we prove it in \cref{sec:gap}.

No analysis that bounds the cost of a BST by the funnel bound can show a competitive ratio of $o(\log\log n)$: by \cref{thm:main-informal}, a BST whose cost is at most $f(n)(\Funnel(X)+m)$ on every permutation has $f(n)=\Omega(\log\log n)$. For the funnel bound this barrier is a direct consequence of \cref{thm:main-informal}; Jiang, Lecomte, Weinstein and Yingchareonthawornchai proved barriers of this kind for analyses by the alternation bound \cite{jlwy24}.

Our lower bound on $\OPT(\pi_k)$ is an explicit set of independent rectangles. All of them have the same orientation: each is spanned by two searches of which the earlier one has the larger key, and contains no other search. Demaine et al.\ proved that a set of pairwise independent rectangles of one orientation in a permutation of $m$ keys is a lower bound on $\OPT-m$ \cite[Lemma~4.5]{dhi+09}.

In the geometric model of \cite{dhi+09}, any BST execution is represented by the set of points $(x,t)$ for the keys $x$ touched at each time $t$, and \cite[Lemma~4.5]{dhi+09} applies to every superset of the search points in which each rectangle of this orientation contains a third point. Because signed Greedy constructs a smallest such superset \cite[Lemma~4.2]{dhi+09}, our set of rectangles also bounds from below the number of points that signed Greedy adds. Signed Greedy works on the plane described above, with each time $t$ as a horizontal row, and goes through the searches in order of time. Fix one of the two orientations, say the rectangles whose lower corner lies to the right. When the algorithm reaches a search $p$, it looks at the points in the earlier rows: the searches and the points that it added before. For every such point $q$ with a larger key than $p$ whose closed rectangle with $p$ contains no point of the earlier rows other than $q$, it adds the point with the key of $q$ in the row of $p$, which is the upper right corner of this rectangle. It adds all these points at once, so afterwards every rectangle spanned by $p$ and an earlier point with a larger key contains a third point. We write $\add(X)$, the right-signed count, for the number of points added in this way over all searches, and $\add_L(X)$, the left-signed count, for the mirror version, which considers the earlier points with a smaller key than $p$ (\cref{sec:irb}).
\begin{corollary}[Informal version of \cref{cor:irb-formal}]
\label{cor:irb-informal}
The permutations $\pi_k$ of \cref{thm:main-informal} satisfy
\begin{align*}
\IRB(\pi_k)=\Omega(m\log\log m),\qquad\add(\pi_k)=\Omega(m\log\log m),\qquad\add_L(\pi_k)=O(m).
\end{align*}
Hence the independent rectangle bound can exceed the funnel bound by a factor $\Omega(\log\log m)$, and the two signed Greedy counts can differ by $\Omega(m\log\log m)$.
\end{corollary}

The first statement refutes the conjecture of Iacono that the two bounds are equivalent \cite[Section~3]{iac13} and answers the question recalled by Guo et al.\ \cite[Section~1.1]{gpsw26} for the funnel bound; the comparison with the bound of Derryberry, Sleator and Wang remains open. The bound on $\add_L(\pi_k)$ holds because each point added by left-signed Greedy on a permutation is the corner of its own empty increasing pair, a pair of searches, the later one with the larger key, whose rectangle contains no other search (\cref{lem:left-greedy} and \eqref{eq:addL-Lambda}). The second statement refutes the conjecture of Lecomte and Weinstein that the two signed counts agree up to an additive $O(m)$ \cite[Section~5]{lw20}. Rotating $\pi_k$ by $90$ degrees moves the lower bound to the left-signed count. This refutes an improvement of a bound of Lecomte and Weinstein on the left-signed count, in terms of empty occurrences of the pattern $3142$, which they note would follow if the funnel bound matched the independent rectangle bound (\cref{sec:rotation}).

\paragraph{The idea in brief.} The funnel bound charges a search only for alternations between its left and right funnels, so the charge is at most twice the size of the left funnel plus one. Large right funnels alone cost nothing. The independent rectangle bound, on the other hand, can count the rectangles of right funnels: a point $q$ in the right funnel of a search $p$ spans with $p$ an empty rectangle whose lower corner lies to the right, and every set of such rectangles that are pairwise independent is a lower bound on $\OPT-m$. Two such rectangles fail to be independent in one configuration only, when one is shifted up and to the right of the other so that each contains a corner of the other in its interior (\cref{lem:staggered}, \cref{fig:config}). So we need sequences with a set of such rectangles, no two of them in this configuration, in which the number of rectangles per search grows, while the left funnels stay small.

We obtain them by composing sequences. Cut the keys into intervals, which we call \emph{strips}. A \emph{pair shuffle} is built from one sequence of strips, the \emph{strip word}, and one sequence of keys of each strip, its \emph{in-strip word}. The strip word says which strip is visited next. A visit to a strip searches two keys of that strip: the key at which the previous visit to this strip stopped (at the first visit, the first letter of its in-strip word), and then the next letter of its in-strip word. For example, if a strip is visited three times and its in-strip word is $a\ b\ c\ d$, the three visits search $a\ b$, then $b\ c$, then $c\ d$. So consecutive visits to a strip overlap in one search, and a sequence of $n$ visits has $2n$ searches, while its pieces, the strip word and the in-strip words, have $2n$ letters plus one for each strip.

This equality of lengths is what allows the composition to be repeated. Call a count \emph{additive} if its value on the composed sequence is the sum of its values on the pieces. Per search, the value of an additive count on the composed sequence is then an average of its values on the pieces (up to the one extra letter per strip, which contributes nothing in our construction), so it stays bounded however often we compose. A count that gains one more unit at every visit grows by $1/2$ per search with every composition. The number of empty increasing pairs, which is the total size of the left funnels, is additive up to an error at descending visits, those whose second search has a smaller key than the first (\cref{prop:pairs}).

Independent sets of rectangles gain. A rectangle of the strip word joins two visits, and in the composed sequence it becomes a rectangle that joins a search of the earlier visit with a search of the later visit. Independent sets of the pieces combine in this way into an independent set of the composed sequence, and there is room for more: a rectangle of the strip word can be used twice, once for each of the two searches of the visit at its upper left corner (its later search), unless this visit descends (\cref{lem:compose,lem:compose-indep}). For the sets that we build, almost every visit is the upper corner of a rectangle of the strip word and few visits descend, so this gives one more rectangle for almost every visit. We start from cyclic scans, which search a fixed number of keys (the base) in increasing order and repeat, and iterate the composition $k$ times; choosing a base that grows with $k$ keeps descending visits rare and gives \cref{thm:main-informal}.

\paragraph{What remains open.} On our sequences the optimum is determined up to a constant factor: \cref{thm:main-informal,prop:tango-informal} give $\OPT(\pi_k)=\Theta(m\log\log m)$, and \cref{cor:irb-informal} gives $\IRB(\pi_k)=\Theta(\OPT(\pi_k))$. So they are consistent with the tightness of the independent rectangle bound and of signed Greedy, which remains open. On $\pi_k$, Tango trees are within a constant factor of $\OPT$, since the proof of \cref{prop:tango-formal} bounds their cost by $O(m\log\log m)$; we do not know the cost of splay trees or of Greedy on $\pi_k$.

\paragraph{Related work.} Three lines of work are close to ours: the geometric view of binary search trees, decompositions of bounds under the composition of sequences, and earlier separations between bounds.

Demaine et al.\ \cite{dhi+09} introduced the geometric view, the independent rectangle bound and signed Greedy, and proved the lower bound for independent rectangles of one orientation that we use \cite[Lemma~4.5]{dhi+09}. What we add on this side is the set of rectangles.

Decompositions of $\OPT$ over blocks of keys are known \cite{cgk+15,ccs23}, as are decomposition theorems that bound the cost of the Greedy algorithm from above \cite{cgk+15,gg19}. Jiang et al.\ \cite[Theorem~1]{jlwy24} proved direct-sum theorems for both of Wilber's bounds: under the composition of a sequence of blocks with sequences inside the blocks, the funnel bound is superadditive and the alternation bound subadditive, up to additive $O(m)$. \Cref{lem:compose} is a statement of this kind for independent rectangles of one orientation under pair shuffles, with one more rectangle per visit as an additional term. This term is of the order of the length, which is the source of the separation. Because the funnel bound is superadditive under such compositions, it can stay bounded per search under repeated nesting only if the pieces are together about as long as the composed sequence. The overlap of the pair shuffle makes them exactly as long, up to the number of strips.

The earlier separations go the other way: the sequences of Lecomte and Weinstein have a funnel bound larger than the alternation bound by a factor $\Theta(\log\log n)$ \cite[Theorem~2]{lw20}, and Jiang et al.\ amplify them by such compositions \cite{jlwy24}. Chalermsook et al.\ \cite{cgk+16} survey the known relations between bounds for binary search trees.

Finally, the split of a sequence with repeated keys into a permutation, which gives the repeated searches of a key distinct new keys, is from Chalermsook et al.\ \cite[Appendix~E]{cgk+15}, who showed that it changes $\OPT$ by at most a constant factor. We use that it keeps independent sets of rectangles (\cref{lem:split}).

%% file: sections/02_overview.tex
\section{Technical overview}
\label{sec:overview}

We build a word $X$ of length $m$, with repeated keys, on which the funnel bound is $O(m)$ and a set of independent rectangles, a lower bound on $\OPT$, has size $\Omega(m\log\log m)$, and then turn it into a permutation. This section explains the construction and the two counts that the proof compares; it uses the names of the formal sections, which give the proofs.

\subsection{Two counts}
\label{sec:overview-counts}

\begin{figure}[t]
\centering
\input{figures/configuration}
\caption{Two empty decreasing pairs, drawn as their rectangles, with keys horizontal and time upwards. (a) The only configuration in which they are not independent, \eqref{eq:ov-staggered}: the second rectangle (dashed) is shifted up and to the right of the first, and each has a corner, marked by a circle, in the interior of the other. (b), (c) Independent pairs: the rectangles may cross, and they may share a point and part of their boundary.}
\label{fig:config}
\end{figure}
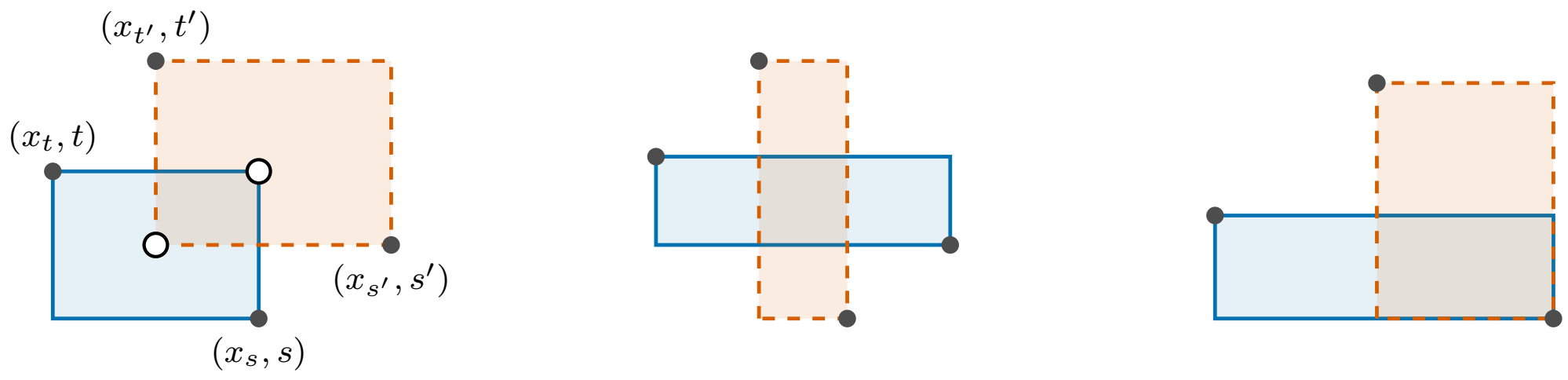

For a point $p$ of $X$, the funnel bound charges the number $f(p)$ of blocks of equal letters in the time-ordered word of its funnel over $\{L,R\}$, where $L$ marks a point of the left funnel $F_L(p)$, the earlier points to the lower left of $p$ whose rectangle with $p$ contains no other search, and $R$ a point of the right funnel. A word with $\ell$ letters $L$ has at most $2\ell+1$ blocks, so only the left funnels matter. Their total size $\Lambda(X)\coloneqq\sum_p\abs{F_L(p)}$ is the number of \emph{empty increasing pairs}, the pairs of searches $s<t$ with $x_s<x_t$ whose rectangle contains no other search. Summing over the $m$ points gives (\cref{lem:funnel-lambda})
\begin{align}
\Funnel(X)&=\sum_pf(p)\notag\\*
&\le\sum_p\bigl(2\abs{F_L(p)}+1\bigr)\notag\\*
&=m+2\Lambda(X).
\label{eq:ov-funnel}
\end{align}
On the other side we use the independent rectangles of Demaine et al.\ \cite{dhi+09}. An \emph{empty decreasing pair} of $X$ is a pair of positions $s<t$ with $x_s>x_t$ whose rectangle $B(s,t)\coloneqq[x_t,x_s]\times[s,t]$ contains no other search; its \emph{lower point} $(x_s,s)$ is a point of the right funnel of its \emph{upper point} $(x_t,t)$. Two such pairs are \emph{independent} if no corner of the rectangle of one lies in the interior of the rectangle of the other. Draw the nodes that a BST touches as points, the node $x$ touched at the $t$-th search as $(x,t)$. These points include the searches, and every rectangle spanned by two of them in different rows and columns contains a third. By \cite{dhi+09} (\cref{sec:prelim-points}), $\OPT(X)$ is the least size of a point set $S$ with these two properties. When $X$ is a permutation of $m$ keys and $F$ is a set of pairwise independent empty decreasing pairs of $X$, every such point set $S$ contains at least $\abs F$ points besides the $m$ searches \cite[Lemma~4.5]{dhi+09}, so (\cref{lem:dhikp})
\begin{align}
\OPT(X)\ge m+\abs{F}.
\label{eq:ov-opt}
\end{align}
Independence is a condition on two rectangles, and it fails in one configuration only. Two empty decreasing pairs $(s,t)$ and $(s',t')$ are independent unless, in one of their two orders (\cref{lem:staggered}),
\begin{align}
x_t<x_{t'}<x_s<x_{s'}\qquad\text{and}\qquad s<s'<t<t',
\label{eq:ov-staggered}
\end{align}
that is, unless the second rectangle is shifted up and to the right of the first and overlaps it, so that each contains a corner of the other in its interior (\cref{fig:config}). Every argument about independence below excludes this configuration.

Finally, the split $X'$ of $X$ gives each search its own key, ordered after the earlier searches of the same key:
\begin{align*}
x'_t\coloneqq(x_t,o_t),\qquad o_t\coloneqq\abs{\{s\le t:x_s=x_t\}},
\end{align*}
with the keys $(x,o)$ ordered by $x$ first and by $o$ second. If $X$ has $n$ distinct keys, then (\cref{lem:split})
\begin{align*}
\Lambda(X')=\Lambda(X)+m-n,
\end{align*}
and a set of independent empty decreasing pairs of $X$ is one of $X'$. So \cref{thm:main-formal} follows from a word $X$ and a set $F$ of pairwise independent empty decreasing pairs of $X$ with
\begin{align}
\Lambda(X)\le2m,\qquad\abs F\ge\frac k2m,\qquad\log_2\log_2m\le2k+2,
\label{eq:ov-goal}
\end{align}
since then
\begin{align*}
\Funnel(X')&\le m+2\Lambda(X')\le m+2(\Lambda(X)+m)\le m+2(2m+m)=7m,\\*
\OPT(X')&\ge m+\abs F\ge\Bigl(1+\frac k2\Bigr)m.
\end{align*}
For the funnel bound, the first inequality is \eqref{eq:ov-funnel} and the second is the split. The bound on the optimum is \eqref{eq:ov-opt} for the permutation $X'$ and the set $F$.

So we need a word whose left funnels stay small and which has a set of independent empty decreasing pairs whose size per search grows. The pairs with a common upper point are always independent of each other, since \eqref{eq:ov-staggered} needs two different upper points. The difficulty is to choose pairs at different searches that do not obstruct each other.

\subsection{Pair shuffles}
\label{sec:overview-shuffle}

We compose words as follows. A \emph{strip word} $U=u_1u_2\cdots$ over a set of strips, with no two equal consecutive letters, decides which strip is visited next: its position $v$ is the $v$-th \emph{visit}, to the strip $u_v$. Each strip $\kappa$ has an \emph{in-strip word} $V_\kappa=V_\kappa[0]V_\kappa[1]\cdots$ over labels, again with no two equal consecutive letters, of length one more than the number of visits to $\kappa$. Keys are pairs $(\kappa,h)$ of a strip and a label, ordered by strip first. The visit $v$, if it is the $i$-th visit to $\kappa=u_v$, searches two keys, at the times $2v-1$ and $2v$:
\begin{align*}
y_{2v-1}\coloneqq(\kappa,V_\kappa[i-1]),\qquad y_{2v}\coloneqq(\kappa,V_\kappa[i]).
\end{align*}
We call these two searches the \emph{first point} and the \emph{second point} of the visit. This word $Y$ is the \emph{pair shuffle} of $U$ and $(V_\kappa)$ (\cref{def:shuffle}). Consecutive visits to a strip \emph{overlap}: the second search of a visit repeats as the first search of the next visit to the same strip. A visit \emph{descends} if $V_\kappa[i]<V_\kappa[i-1]$. Otherwise $V_\kappa[i]>V_\kappa[i-1]$, and we also say that it ascends.

Our running example has three strips and three labels:
\begin{align*}
U=(0\ 1\ 2)^3,\qquad V_0=V_1=V_2=0\ 1\ 2\ 0.
\end{align*}
Writing $\kappa_h$ for the key $(\kappa,h)$ with strip $\kappa$ and label $h$, its pair shuffle is the word of $18$ searches
\begin{align*}
Y=0_00_1\ 1_01_1\ 2_02_1\quad0_10_2\ 1_11_2\ 2_12_2\quad0_20_0\ 1_21_0\ 2_22_0,
\end{align*}
grouped into its nine visits; the last three visits descend (drawn as the dots and arrows in \cref{fig:example} below). Restricted to a strip and with consecutive repeats merged, $Y$ gives back $V_\kappa$, and with each visit merged to its strip, $Y$ gives back $U$. Because of the overlap, the pieces together are as long as $Y$, up to the number $q$ of strips, as in \eqref{eq:lengths}:
\begin{align}
\abs{Y}=2\abs{U},\qquad\sum_\kappa\abs{V_\kappa}=\abs{U}+q.
\label{eq:ov-lengths}
\end{align}
This is the property that the whole construction rests on. If a count $c$ is additive under pair shuffles, then its value per search is an average of the values per search of the pieces, with weights $1/2$ and $1/2$, and a gain of one per visit adds $1/2$ per search:
\begin{align*}
\frac{c(U)+\sum_\kappa c(V_\kappa)}{\abs Y}=\frac12\cdot\frac{c(U)}{\abs U}+\frac12\cdot\frac{\sum_\kappa c(V_\kappa)}{\sum_\kappa\abs{V_\kappa}-q},\qquad\frac{\abs U}{\abs Y}=\frac12.
\end{align*}
In our construction each $V_\kappa$ is formed by appending one letter with zero contribution to a word $V$ of length $\abs U/q$, so the second denominator $\sum_\kappa\abs{V_\kappa}-q=q\abs V$ is the exact total length of those words $V$. Thus an additive count stays bounded per search when we iterate, while a count that gains a constant per visit grows by a constant per iteration.

\subsection{The left funnels are additive}
\label{sec:overview-pairs}

An empty increasing pair inside a strip is an empty increasing pair of its in-strip word. Each position of $V_\kappa$ other than the first and the last appears twice in a row in the strip, as the second point of one visit and the first point of the next, so two positions give up to four pairs of copies; only the later copy of the earlier position with the earlier copy of the later position can be empty, because each other pair has a copy inside its rectangle.

An empty increasing pair across two strips $\kappa_1<\kappa_2$, from a visit $v_1$ to a visit $v_2>v_1$, needs that no visit in between goes to a strip strictly between them:
\begin{align*}
u_w\notin(\kappa_1,\kappa_2)\qquad(v_1<w<v_2).
\end{align*}
Consider such a pair whose earlier point is not the first point of a descending visit and whose later point is not the second point of one. If the later point were the second point of $v_2$, then $v_2$ would ascend, so its first point would lie in the rectangle; hence the later point is the first point of $v_2$. By the overlap, the last visit to $\kappa_2$ before $v_2$, if there is one, ended with the same key, and a search of this key between the two points of the pair would lie in their rectangle; so that visit comes before $v_1$. Symmetrically, the earlier point is the second point of $v_1$ and $\kappa_1$ is not visited again before $v_2$. Thus the pair joins the second point of $v_1$ with the first point of $v_2$, and
\begin{align*}
u_w\notin[\kappa_1,\kappa_2]\qquad(v_1<w<v_2),
\end{align*}
which says that $(v_1,v_2)$ is an empty increasing pair of $U$. Conversely, every empty increasing pair of $U$ arises in this way. So (\cref{prop:pairs})
\begin{align}
\Lambda(Y)=\Lambda(U)+\sum_\kappa\Lambda(V_\kappa)+\err(Y),
\label{eq:ov-pairs}
\end{align}
where $\err(Y)$ counts the remaining empty increasing pairs across strips: those that start at the first point of a descending visit or end at the second point of one. In the example, $\Lambda(U)=6$, $\Lambda(V_\kappa)=2$, and the six pairs counted by $\err(Y)$ all involve the three descending visits at the end:
\begin{align*}
\Lambda(Y)=18=6+3\cdot2+6.
\end{align*}

To bound $\err(Y)$, let $A(Y)$ be the largest number of empty increasing pairs that have a common later point, or a common earlier point. A visit to $\kappa$ descends at a descent of $V_\kappa$, a position where the next label is smaller, so with $\desc(V_\kappa)$ the number of descents there are $\sum_\kappa\desc(V_\kappa)$ descending visits, and \eqref{eq:err-degree}
\begin{align*}
\err(Y)\le2A(Y)\cdot\sum_\kappa\desc(V_\kappa).
\end{align*}
The words that we build in \cref{sec:overview-iterate} keep $A$ small, and we bound it there.

\subsection{One more rectangle per visit}
\label{sec:overview-rect}

Let $F_U$ be a set of independent empty decreasing pairs of the strip word and $F_\kappa$ one of each in-strip word. Each of these pairs has a copy in $Y$. A pair $(i',i)$ of positions of $V_\kappa$ is copied to the later copy of the position $i'$ and the earlier copy of the position $i$, that is, to the first point of the $(i'+1)$-th visit to $\kappa$ and the second point of the $i$-th visit. A pair $(w,v)$ of $F_U$ is copied to the second point of the visit $w$ and the first point of the visit $v$, the pair of times
\begin{align*}
(2w,\ 2v-1).
\end{align*}
Both copies are empty decreasing pairs of $Y$. For the first, the points of the strip between them in time are copies of the positions between $i'$ and $i$. For the second, the points between them belong to the visits between $w$ and $v$, and these go to strips outside $[u_v,u_w]$, because $(w,v)$ is an empty decreasing pair of $U$. The gain comes from the second point of $v$. If $v$ does not descend, its second point has a larger key than its first, so the first point does not lie in the rectangle of
\begin{align*}
(2w,\ 2v),
\end{align*}
which is therefore an empty decreasing pair as well: the pair $(w,v)$ can be used a second time. We do this once for every visit $v$ that is an upper point of $F_U$ and does not descend, choosing $w=w_v$ to be the least position such that $(w,v)\in F_U$. At every descending visit $v$, its two points $(2v-1,2v)$ are consecutive in time with $y_{2v-1}>y_{2v}$, so they also form an empty decreasing pair (which may already be one of the in-strip copies). Let $F$ be the set of all copies, second uses, and pairs $(2v-1,2v)$ at descending visits.

\begin{figure}[t]
\centering
\input{figures/example}

\smallskip
{\scriptsize\legendbox{rst}\ copy of a pair of $F_U$\qquad\legendbox{rplus}\ second use\qquad\legendbox{rin}\ copy of a pair of $F_\kappa$}
\caption{The running example. The dots are the $18$ searches of $Y$, with keys horizontal and time upwards. An arrow joins the first point of every visit to its second point, and points to the left in the three descending visits. The nine rectangles are the pairs of $F$. Solid: the copies of the four pairs of $F_U$. Dashed, with shared bottom and right edges drawn slightly inside so that they stay visible: the second uses at the visits $4$ and $5$. Dotted: the copies of the pairs of the $F_\kappa$, which coincide with the pairs $(2v-1,2v)$ of the descending visits $7$, $8$ and $9$. The four rectangles with lower point $2_1$ at time $6$ share this corner, and no two of the nine rectangles are in the configuration of \cref{fig:config}(a).}
\label{fig:example}
\end{figure}
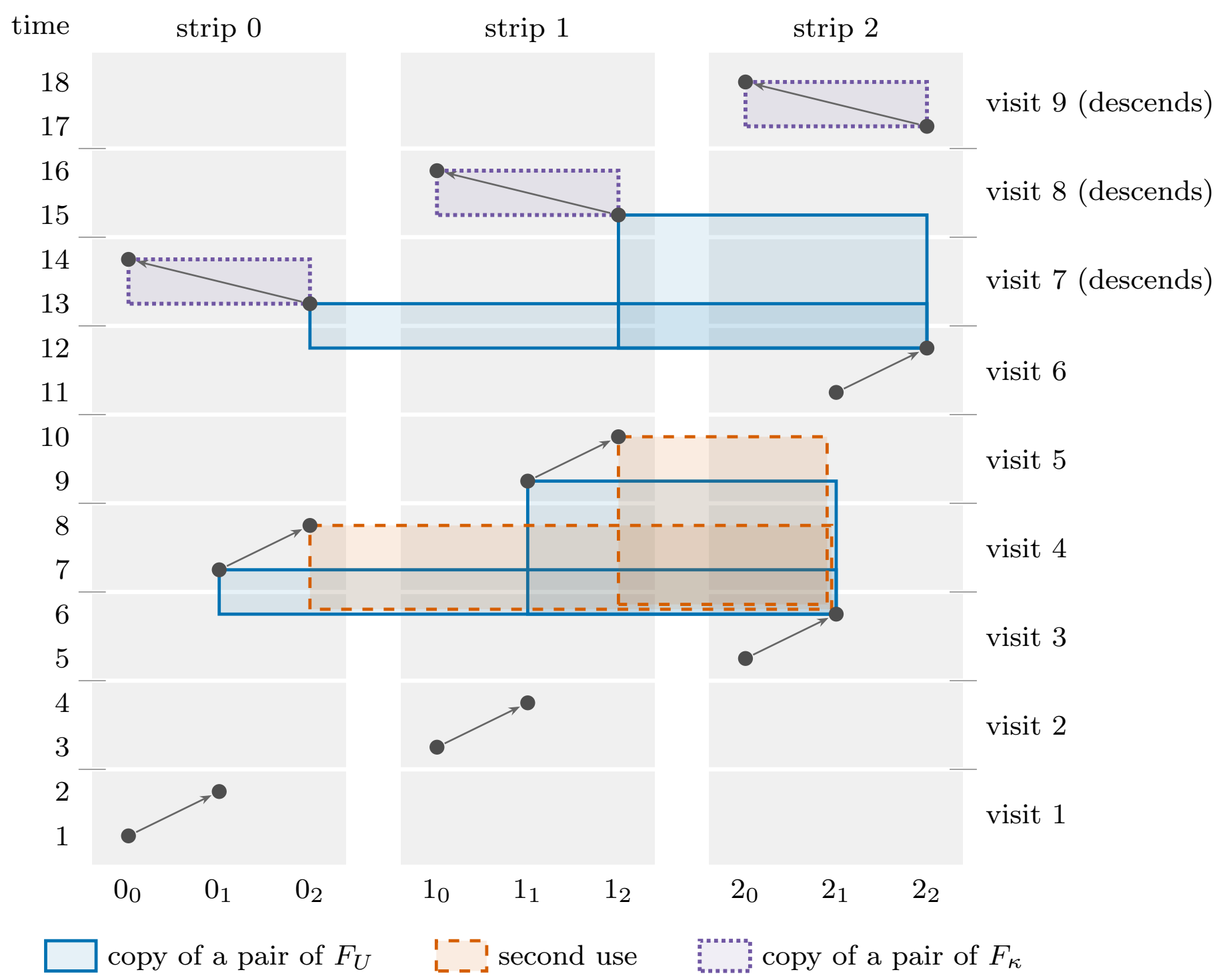

In the running example, take for $F_U$ the pairs $(3,4),(3,5),(6,7),(6,8)$ of $U$, which join the visits to the strips $0$ and $1$ in the second and third repetitions of $0\ 1\ 2$ with the preceding visit to the strip $2$, and for every strip the pair of the last two letters $2\ 0$ of $V_\kappa$. Here the pairs $(2v-1,2v)$ at the three descending visits $7$, $8$ and $9$ are already the copies of the $F_\kappa$, so $F$ consists of nine pairs of times (\cref{fig:example}):
\begin{align*}
\underbrace{(6,7),\ (6,9),\ (12,13),\ (12,15)}_{\text{copies of }F_U},\qquad\underbrace{(6,8),\ (6,10)}_{\text{second uses}},\qquad\underbrace{(13,14),\ (15,16),\ (17,18)}_{\text{copies of the }F_\kappa}.
\end{align*}
The second uses are at the visits $4$ and $5$, whose second points $0_2$ and $1_2$, at the times $8$ and $10$, are joined with the search of $2_1$ at time $6$. The visits $7$ and $8$ are upper points of $F_U$ as well, but they descend.

The set $F$ is independent (\cref{lem:compose-indep}). A pair $(2v-1,2v)$ at a descending visit joins consecutive times, whereas \eqref{eq:ov-staggered} requires a time strictly between the two times of each pair, and a copy of a pair of an in-strip word is kept apart from the other pairs by the strips. The real case is two pairs that come from pairs $(w_1,v_1)$ and $(w_2,v_2)$ of $F_U$, with the rectangle of the second shifted up and to the right as in \eqref{eq:ov-staggered}. Ordering times and keys (by strip first) in \eqref{eq:ov-staggered} gives
\begin{align*}
w_1<w_2<v_1\le v_2,\qquad u_{v_1}\le u_{v_2}\le u_{w_1}\le u_{w_2}.
\end{align*}
If $v_1<v_2$, the emptiness of $(w_1,v_1)$ at $w_2$ and of $(w_2,v_2)$ at $v_1$ makes the two outer strip inequalities strict, and if $u_{v_2}<u_{w_1}$ as well, the two pairs satisfy \eqref{eq:ov-staggered} in $U$, which the independence of $F_U$ excludes. What remains is $u_{v_2}=u_{w_1}$, and again by emptiness no visit between $w_1$ and $v_2$ goes to this strip. Here the overlap decides: the first point of $v_2$ repeats the key of the second point of $w_1$,
\begin{align*}
y_{2v_2-1}=y_{2w_1},
\end{align*}
and any second use at $v_2$ has $y_{2v_2}>y_{2v_2-1}$, so the upper point of the second pair does not lie to the left of the lower point of the first, as \eqref{eq:ov-staggered} would require (\cref{fig:two-strip}). If $v_1=v_2$, the two pairs are the copy of $(w_1,v_1)$ and the second use at the same visit, of a pair $(w_2,v_1)$ with $w_2>w_1$, and the choice of the least position $w_{v_1}$ excludes this.

Counting only the disjoint copies and second uses, which exist at all upper points of $F_U$ except the at most $\sum_\kappa\desc(V_\kappa)$ descending ones, gives (\cref{lem:compose})
\begin{align}
\abs F\ge\abs{F_U}+\sum_\kappa\abs{F_\kappa}+\abs{\tops(F_U)}-\sum_\kappa\desc(V_\kappa),\qquad\abs{\tops(F)}\ge2\abs{\tops(F_U)},
\label{eq:ov-rect}
\end{align}
where $\tops(F)$ is the set of the upper points of $F$. The second bound holds because both points of every visit $v\in\tops(F_U)$ are upper points of $F$: $2v-1$ by the copy of a pair ending at $v$, and $2v$ either by the second use at $v$ or, when $v$ descends, by $(2v-1,2v)$. Keeping the share of upper points high is what allows the next level to gain again. In the running example, $\abs F=9$, while the first bound of \eqref{eq:ov-rect} gives $4+3+4-3=8$. One level of nesting does not yet separate the two counts, since $\Lambda(Y)=18$. The separation comes from repeating the composition, because each level adds a constant per search to $\abs F$ and almost nothing per search to $\Lambda$ (\cref{sec:overview-iterate}).

\subsection{Iterating}
\label{sec:overview-iterate}

We nest pair shuffles, starting from cyclic scans. A \emph{level-$0$ word} is a cyclic scan
\begin{align*}
W_0[r]\coloneqq(0\ 1\ \cdots\ g-1)^r,
\end{align*}
the keys $0,\dots,g-1$ in increasing order, repeated for $r$ \emph{rounds}; the number $g$ of keys is the \emph{base}. A level-$1$ word is a pair shuffle of two cyclic scans. Its strip word $U$ is a cyclic scan, so the keys of $U$ become the strips, and every strip gets the same in-strip word $V'$, a second cyclic scan $V$ followed by its first letter:
\begin{align*}
U=W_0[r_U],\qquad V=W_0[r],\qquad V'=V\ V[0].
\end{align*}
A pair shuffle needs an in-strip word with one more letter than the strip has visits. The appended letter $V[0]$ supplies it, provided every strip is visited exactly $\abs V$ times, that is, $r_U=\abs V=gr$. The running example is this word for $g=3$ and $r=1$: the strip word $(0\ 1\ 2)^3$ has $r_U=3$ rounds, and $V'=0\ 1\ 2\ 0$.

Level $j$ repeats this step with level-$(j-1)$ words in place of the two cyclic scans. The level-$j$ word $W_j[r]$ is the pair shuffle of a level-$(j-1)$ strip word $U$, whose keys become the strips, with the same in-strip word $V'=V\ V[0]$ for every strip, where the \emph{label word} $V$ is a level-$(j-1)$ word as well (\cref{def:level}):
\begin{align*}
U\coloneqq W_{j-1}[g^{2^{j-1}}r],\qquad V\coloneqq W_{j-1}[r].
\end{align*}
Every key of $U$ must occur exactly $\abs V$ times in $U$, so the cyclic scans inside the strip word have more rounds than those inside the label word. The counts fit, because the length of $W_j[r]$ and the number of occurrences of each of its keys are
\begin{align*}
\abs{W_j[r]}=2^jg^{2^j}r\qquad\text{and}\qquad2^jr,
\end{align*}
so every key of $U=W_{j-1}[g^{2^{j-1}}r]$ occurs $2^{j-1}g^{2^{j-1}}r=\abs{W_{j-1}[r]}=\abs V$ times. Moreover $W_j[r]$ is the period $W_j[1]$ repeated $r$ times, which we again call its rounds, and its first letter is its smallest key (\cref{lem:level-basic}). The running example is $W_1[1]$ with $g=3$.

The level words keep the error of \eqref{eq:ov-pairs} small, because few empty increasing pairs meet at one point. Let $p\coloneqq(x,t)$ be a point of $W\coloneqq W_j[r]=w_1\cdots w_m$, and let $(w_s,s)$ be the earlier point of an empty increasing pair with $p$. Then
\begin{align*}
w_s<x,\qquad w_{s'}\notin[w_s,x]\quad(s<s'<t).
\end{align*}
Because $W$ repeats its period $P_j\coloneqq W_j[1]$ of length $d_j\coloneqq2^jg^{2^j}$ and $P_j$ contains every key, $w_s$ would be searched between $s$ and $t$ if $t-s>d_j$; thus $s$ lies in the \emph{factor} (contiguous subword) of length at most $d_j$ immediately before $t$. We cannot bound these pairs directly by induction on $j$: when we project to the strip word $U$, a later visit before $t$ can return to the strip $a$ of $w_s$ and search keys of $a$ below $w_s$, so in $U$ the letter $a$ can repeat after the visit of $s$, even though no strip strictly between $a$ and the strip just below $x$ is visited. We therefore relax the closed interval $[w_s,x]$ to the open interval $(w_s,x)$, allowing later searches of the same key. Formally, for a word $Z=z_1\cdots z_l$, a position $s$ is a \emph{lower record} of $Z$ below $x$ if
\begin{align*}
z_s<x,\qquad z_{s'}\notin(z_s,x)\quad(s<s'\le l).
\end{align*}
For example, in $Z=2\ 1\ 2$ with $x=3$, both searches of $2$ (positions $1$ and $3$) are lower records, whereas the search of $1$ is disqualified by the later $2\in(1,3)$. Since $(w_s,x)\subset[w_s,x]$, every earlier point of an empty increasing pair with $p$ is a lower record below $x$ of the factor of length at most $d_j$ before $t$. Upper records above $x$ are defined symmetrically, with time and the key order reversed, and contain all later points of empty increasing pairs starting at $p$. Let $H_j$ be the maximum number of lower records below $x$, or of upper records above $x$, over all $r\ge1$, all factors $Z$ of $W_j[r]$ of length at most $d_j$, and all thresholds $x$ (including $\pm\infty$). Then $A(W_j[r])\le H_j$.

To bound $H_j$, consider a factor $Z$ of length $d_j$ (one full period) of $W$, let $x$ lie in strip $c$, and let $b$ be the strip immediately below $c$; since $Z$ has length $d_j$, it visits every strip. Every lower record below $x$ lies either in a strip $a<b$, in $b$, or in $c$. Suppose a lower record $s$ lies in a strip $a<b$, at visit $v$. Every key of $b$ lies strictly between $w_s$ and $x$, so $s$ comes after the last visit to $b$ in $Z$. Because every period of $U$ visits $b$, the visits in $Z$ after that last visit to $b$ form a factor of $U$ of length less than $d_{j-1}$. Each visit $w>v$ in this factor has a point after $s$ whose key avoids $(w_s,x)$, so its strip avoids $(a,b)$:
\begin{align*}
u_w\notin(a,b)\qquad(w>v).
\end{align*}
Thus $v$ is a lower record below $b$ of this factor of $U$ (later visits to $a$ itself do not violate $u_w\notin(a,b)$, which is why lower records allow ties). Since each visit has two points, all strips below $b$ together hold at most $2H_{j-1}$ lower records.

The lower records in strip $b$ or in strip $c$ are lower records among the labels searched in that strip. Across consecutive visits to one strip, the labels of the first points form a factor of $V'=W_{j-1}[r]\ V[0]$, which is a prefix of $W_{j-1}[r+1]$ because $V[0]$ is the first letter of the period, and the same holds for the second points; since one period of $W$ makes $d_j/2=qd_{j-1}$ visits in total and visits every strip equally often, it visits each strip $d_{j-1}$ times, so each of these label factors has length at most $d_{j-1}$ (\cref{sec:level-pairs}). Hence each of the strips $b$ and $c$ holds at most $2H_{j-1}$ lower records ($H_{j-1}$ at first points and $H_{j-1}$ at second points). Adding the three contributions (\cref{fig:records}) gives
\begin{align*}
H_j\le\underbrace{2H_{j-1}}_{\text{strips below }b}+\underbrace{2\cdot2H_{j-1}}_{\text{strips }b\text{ and }c}=6H_{j-1}.
\end{align*}
For factors $Z$ shorter than one period, which need not visit $b$, the formal proof (\cref{sec:level-pairs}) replaces $b$ by the largest strip below $c$ that $Z$ does visit. At level $0$, a factor of length at most $g$ of a cyclic scan consists of at most two increasing runs, each contributing at most one record, so $H_0\le2$. Upper records satisfy the same bound by reversal symmetry, giving (\cref{lem:pairs-records,lem:records-recursion})
\begin{align*}
A\bigl(W_j[r]\bigr)\le H_j\le2\cdot6^j,
\end{align*}
so the error per descending visit grows by at most a factor $6$ per level.

On a cyclic scan, the left funnel of a search of $i>0$ is the previous search, and from the second round on, every search of a key below $g-1$ forms an empty decreasing pair with the preceding search of $g-1$. These pairs have a common key at their lower points, so no two of them satisfy \eqref{eq:ov-staggered}. So a cyclic scan has about one empty increasing pair per search, and these independent empty decreasing pairs, which we collect in $F$, give about one pair and one upper point per search:
\begin{align*}
\Lambda(W_0[r])=r(g-1),\qquad\abs F=\abs{\tops(F)}=(r-1)(g-1),\qquad\abs{W_0[r]}=rg.
\end{align*}
At level $1$, for $W_1[1]$, with $U=W_0[g]$ and $V=0\ 1\ \cdots\ g-1$, the set of $U$ has $(g-1)^2$ pairs and as many upper points, we take no pairs of $V'$, and one visit per strip descends. On the $2g^2$ searches, \eqref{eq:ov-rect} and \eqref{eq:ov-pairs} give
\begin{align*}
\abs F\ge(g-1)^2+(g-1)^2-g=2g^2-5g+2,\qquad\Lambda(Y)=2g(g-1)+\err(Y),
\end{align*}
with $\err(Y)\le2\cdot12\cdot g$, since $g$ visits descend and every point lies in at most $H_1\le12$ empty increasing pairs of each kind. Per search, the set gains about $1/2$ over the average of its pieces, while the empty increasing pairs stay at about one.

For general $j$, divide \eqref{eq:ov-pairs} and \eqref{eq:ov-rect} by $\abs{Y}$ and use \eqref{eq:ov-lengths}. Call the steps between consecutive letters, together with the step from the last letter back to the first, the \emph{cyclic steps} of a word. Every level word has cyclic-descent fraction $1/g$---a cyclic scan has one descent per round of length $g$, and each higher level averages the fractions of its strip and label words (\cref{lem:level-basic})---so $V'=V\ V[0]$ has $\abs V/g$ descents and $Y=W_j[r]$ has $q\abs V/g=\abs Y/(2g)$ descending visits. The appended letter $V[0]$ is the smallest key, so it ends no empty increasing pair and $\Lambda(V')=\Lambda(V)$. Write $\lambda(W)\coloneqq\Lambda(W)/\abs W$ and, for the set $F$ that \eqref{eq:ov-rect} builds recursively on level words (taking $F_\kappa$ on $V'=V\ V[0]$ to be the set built for its prefix $V$), write $\alpha(W)\coloneqq\abs F/\abs W$ and $\gamma(W)\coloneqq\abs{\tops(F)}/\abs W$. With the strip word $U$ and the label word $V$ of level $j-1$, the proofs of \cref{prop:lambda-level,prop:rect-level} give
\begin{align}
\lambda(Y)&\le\frac{\lambda(U)+\lambda(V)}2+\frac{2\cdot6^j}g,\notag\\
\alpha(Y)&\ge\frac{\alpha(U)+\alpha(V)}2+\frac{\gamma(U)}2-\frac1{2g},\notag\\
\gamma(Y)&\ge\gamma(U).
\label{eq:ov-recursions}
\end{align}
In the first line, the $\abs Y/(2g)$ descending visits add at most $2A(Y)\cdot\abs Y/(2g)\le(2\cdot6^j/g)\abs Y$ pairs to $\err(Y)$, or $2\cdot6^j/g$ per search. In the second line, $\gamma(U)/2$ counts the second uses, one for every upper point of the set of $U$, and $1/(2g)$ subtracts the descending visits, which have none. The share $\gamma$ of upper points never decreases from a strip word to the composed word. Following strip words down, we reach a cyclic scan with $r'\ge g$ rounds, since the strip word of $W_j[r]$ has $g^{2^{j-1}}r\ge g$ rounds and the number of rounds grows along strip words; its share of upper points is $(1-1/g)(1-1/r')$. So $\gamma(U)\ge(1-1/g)^2$, and each level adds at least
\begin{align*}
\frac{(1-1/g)^2}2-\frac1{2g}\ge\frac12\Bigl(1-\frac3g\Bigr)
\end{align*}
to the average of $\alpha$ over the pieces. At the bottom of the recursion are cyclic scans with $r'\ge r$ rounds, whose share $\alpha=(1-1/g)(1-1/r')$ is at least $\tfrac34(1-1/g)$ for $r=4$. So the level-$k$ word $W$ with $r=4$ has
\begin{align*}
\alpha(W)\ge\frac34\Bigl(1-\frac1g\Bigr)+\frac k2\Bigl(1-\frac3g\Bigr)=\frac k2+\frac34\Bigl(1-\frac{2k+1}g\Bigr),
\end{align*}
which is at least $k/2$ when $g\ge2k+1$. Taking the base $g=3\cdot6^k$, large enough to absorb the error terms of all $k$ levels, and $r=4$ gives the level-$k$ word $W$ with
\begin{align*}
\lambda(W)\le1+\underbrace{\frac2g\sum_{j=1}^k6^j}_{\text{descending visits}}<2,\qquad\alpha(W)\ge\underbrace{\frac k2}_{\text{second uses}},\qquad m=2^{k+2}g^{2^k},
\end{align*}
and, since $\log_2g\le3k+2$,
\begin{align*}
\log_2m=k+2+2^k\log_2g\le2^{2k+2}.
\end{align*}
This is \eqref{eq:ov-goal}.

Our bounds need a base that grows with the number of levels, since the bound on the extra pairs per descending visit grows by a factor $6$ per level; a base exponential in $k$ costs only a lower-order term, since $\log_2\log_2m=k+O(\log k)$. The weights $1/2$ and $1/2$ in \eqref{eq:ov-recursions} come from \eqref{eq:ov-lengths}, that is, from the overlap. If a visit searched two fresh labels instead, the in-strip words alone would be as long as $Y$, and the weights would be $1/2$ and $1$. Then the first recursion would let $\lambda$ grow by a constant factor at every level, and the argument would give no separation.

%% file: figures/configuration.tex
\begin{tikzpicture}[x=0.62cm,y=0.62cm]
\useasboundingbox (-11.2,-0.75) rectangle (11.2,5.35);

\draw[rst] (-10.2,1) rectangle (-7.4,3);
\draw[rplus] (-8.8,2) rectangle (-5.6,4.5);
\node[sdot] at (-7.4,1) {}; \node[sdot] at (-10.2,3) {};
\node[sdot] at (-5.6,2) {}; \node[sdot] at (-8.8,4.5) {};
\node[cdot] at (-8.8,2) {}; \node[cdot] at (-7.4,3) {};
\node[figtxt,anchor=north] at (-7.4,0.82) {$(x_s,s)$};
\node[figtxt,anchor=south] at (-10.2,3.16) {$(x_t,t)$};
\node[figtxt,anchor=north] at (-5.6,1.82) {$(x_{s'},s')$};
\node[figtxt,anchor=south] at (-8.8,4.66) {$(x_{t'},t')$};
\node[figtxt,anchor=base] at (-7.9,-0.45) {(a) not independent};

\draw[rst] (-2.0,2) rectangle (2.0,3.2);
\draw[rplus] (-0.6,1) rectangle (0.6,4.5);
\node[sdot] at (2.0,2) {}; \node[sdot] at (-2.0,3.2) {};
\node[sdot] at (0.6,1) {}; \node[sdot] at (-0.6,4.5) {};
\node[figtxt,anchor=base] at (0.0,-0.45) {(b) independent};

\draw[rst] (5.6,1) rectangle (10.2,2.4);
\draw[rplus] (7.8,1) rectangle (10.2,4.2);
\node[sdot] at (10.2,1) {}; \node[sdot] at (5.6,2.4) {}; \node[sdot] at (7.8,4.2) {};
\node[figtxt,anchor=base] at (7.9,-0.45) {(c) independent};
\end{tikzpicture}

%% file: figures/example.tex
\begin{tikzpicture}[x=0.90cm,y=0.44cm]
\fill[black!6] (-0.40,0.35) rectangle (2.40,18.65);
\draw[white,line width=1.2pt] (-0.42,2.5) -- (2.42,2.5);
\draw[white,line width=1.2pt] (-0.42,4.5) -- (2.42,4.5);
\draw[white,line width=1.2pt] (-0.42,6.5) -- (2.42,6.5);
\draw[white,line width=1.2pt] (-0.42,8.5) -- (2.42,8.5);
\draw[white,line width=1.2pt] (-0.42,10.5) -- (2.42,10.5);
\draw[white,line width=1.2pt] (-0.42,12.5) -- (2.42,12.5);
\draw[white,line width=1.2pt] (-0.42,14.5) -- (2.42,14.5);
\draw[white,line width=1.2pt] (-0.42,16.5) -- (2.42,16.5);
\node[figtxt,anchor=south] at (1.00,18.8) {strip $0$};
\node[figtxt,anchor=north] at (0.00,0.1) {$0_0$};
\node[figtxt,anchor=north] at (1.00,0.1) {$0_1$};
\node[figtxt,anchor=north] at (2.00,0.1) {$0_2$};
\fill[black!6] (3.00,0.35) rectangle (5.80,18.65);
\draw[white,line width=1.2pt] (2.98,2.5) -- (5.82,2.5);
\draw[white,line width=1.2pt] (2.98,4.5) -- (5.82,4.5);
\draw[white,line width=1.2pt] (2.98,6.5) -- (5.82,6.5);
\draw[white,line width=1.2pt] (2.98,8.5) -- (5.82,8.5);
\draw[white,line width=1.2pt] (2.98,10.5) -- (5.82,10.5);
\draw[white,line width=1.2pt] (2.98,12.5) -- (5.82,12.5);
\draw[white,line width=1.2pt] (2.98,14.5) -- (5.82,14.5);
\draw[white,line width=1.2pt] (2.98,16.5) -- (5.82,16.5);
\node[figtxt,anchor=south] at (4.40,18.8) {strip $1$};
\node[figtxt,anchor=north] at (3.40,0.1) {$1_0$};
\node[figtxt,anchor=north] at (4.40,0.1) {$1_1$};
\node[figtxt,anchor=north] at (5.40,0.1) {$1_2$};
\fill[black!6] (6.40,0.35) rectangle (9.20,18.65);
\draw[white,line width=1.2pt] (6.38,2.5) -- (9.22,2.5);
\draw[white,line width=1.2pt] (6.38,4.5) -- (9.22,4.5);
\draw[white,line width=1.2pt] (6.38,6.5) -- (9.22,6.5);
\draw[white,line width=1.2pt] (6.38,8.5) -- (9.22,8.5);
\draw[white,line width=1.2pt] (6.38,10.5) -- (9.22,10.5);
\draw[white,line width=1.2pt] (6.38,12.5) -- (9.22,12.5);
\draw[white,line width=1.2pt] (6.38,14.5) -- (9.22,14.5);
\draw[white,line width=1.2pt] (6.38,16.5) -- (9.22,16.5);
\node[figtxt,anchor=south] at (7.80,18.8) {strip $2$};
\node[figtxt,anchor=north] at (6.80,0.1) {$2_0$};
\node[figtxt,anchor=north] at (7.80,0.1) {$2_1$};
\node[figtxt,anchor=north] at (8.80,0.1) {$2_2$};
\draw[black!45,line width=0.3pt] (-0.55,2.5) -- (-0.25,2.5);
\draw[black!45,line width=0.3pt] (9.05,2.5) -- (9.35,2.5);
\draw[black!45,line width=0.3pt] (-0.55,4.5) -- (-0.25,4.5);
\draw[black!45,line width=0.3pt] (9.05,4.5) -- (9.35,4.5);
\draw[black!45,line width=0.3pt] (-0.55,6.5) -- (-0.25,6.5);
\draw[black!45,line width=0.3pt] (9.05,6.5) -- (9.35,6.5);
\draw[black!45,line width=0.3pt] (-0.55,8.5) -- (-0.25,8.5);
\draw[black!45,line width=0.3pt] (9.05,8.5) -- (9.35,8.5);
\draw[black!45,line width=0.3pt] (-0.55,10.5) -- (-0.25,10.5);
\draw[black!45,line width=0.3pt] (9.05,10.5) -- (9.35,10.5);
\draw[black!45,line width=0.3pt] (-0.55,12.5) -- (-0.25,12.5);
\draw[black!45,line width=0.3pt] (9.05,12.5) -- (9.35,12.5);
\draw[black!45,line width=0.3pt] (-0.55,14.5) -- (-0.25,14.5);
\draw[black!45,line width=0.3pt] (9.05,14.5) -- (9.35,14.5);
\draw[black!45,line width=0.3pt] (-0.55,16.5) -- (-0.25,16.5);
\draw[black!45,line width=0.3pt] (9.05,16.5) -- (9.35,16.5);
\node[figtxt,anchor=east] at (-0.6,1) {$1$};
\node[figtxt,anchor=east] at (-0.6,2) {$2$};
\node[figtxt,anchor=east] at (-0.6,3) {$3$};
\node[figtxt,anchor=east] at (-0.6,4) {$4$};
\node[figtxt,anchor=east] at (-0.6,5) {$5$};
\node[figtxt,anchor=east] at (-0.6,6) {$6$};
\node[figtxt,anchor=east] at (-0.6,7) {$7$};
\node[figtxt,anchor=east] at (-0.6,8) {$8$};
\node[figtxt,anchor=east] at (-0.6,9) {$9$};
\node[figtxt,anchor=east] at (-0.6,10) {$10$};
\node[figtxt,anchor=east] at (-0.6,11) {$11$};
\node[figtxt,anchor=east] at (-0.6,12) {$12$};
\node[figtxt,anchor=east] at (-0.6,13) {$13$};
\node[figtxt,anchor=east] at (-0.6,14) {$14$};
\node[figtxt,anchor=east] at (-0.6,15) {$15$};
\node[figtxt,anchor=east] at (-0.6,16) {$16$};
\node[figtxt,anchor=east] at (-0.6,17) {$17$};
\node[figtxt,anchor=east] at (-0.6,18) {$18$};
\node[figtxt,anchor=east] at (-0.6,19.3) {time};
\node[figtxt,anchor=west] at (9.40,1.5) {visit $1$};
\node[figtxt,anchor=west] at (9.40,3.5) {visit $2$};
\node[figtxt,anchor=west] at (9.40,5.5) {visit $3$};
\node[figtxt,anchor=west] at (9.40,7.5) {visit $4$};
\node[figtxt,anchor=west] at (9.40,9.5) {visit $5$};
\node[figtxt,anchor=west] at (9.40,11.5) {visit $6$};
\node[figtxt,anchor=west] at (9.40,13.5) {visit $7$ (descends)};
\node[figtxt,anchor=west] at (9.40,15.5) {visit $8$ (descends)};
\node[figtxt,anchor=west] at (9.40,17.5) {visit $9$ (descends)};
\draw[rst] (1.00,6.00) rectangle (7.80,7.00);
\draw[rst] (4.40,6.00) rectangle (7.80,9.00);
\draw[rst] (2.00,12.00) rectangle (8.80,13.00);
\draw[rst] (5.40,12.00) rectangle (8.80,15.00);
\draw[rplus] (2.00,6.11) rectangle (7.75,8.00);
\draw[rplus] (5.40,6.22) rectangle (7.70,10.00);
\draw[rin] (0.00,13.00) rectangle (2.00,14.00);
\draw[rin] (3.40,15.00) rectangle (5.40,16.00);
\draw[rin] (6.80,17.00) rectangle (8.80,18.00);
\draw[varrow] (0.00,1) -- (1.00,2);
\draw[varrow] (3.40,3) -- (4.40,4);
\draw[varrow] (6.80,5) -- (7.80,6);
\draw[varrow] (1.00,7) -- (2.00,8);
\draw[varrow] (4.40,9) -- (5.40,10);
\draw[varrow] (7.80,11) -- (8.80,12);
\draw[varrow] (2.00,13) -- (0.00,14);
\draw[varrow] (5.40,15) -- (3.40,16);
\draw[varrow] (8.80,17) -- (6.80,18);
\node[sdot] at (0.00,1) {};
\node[sdot] at (1.00,2) {};
\node[sdot] at (3.40,3) {};
\node[sdot] at (4.40,4) {};
\node[sdot] at (6.80,5) {};
\node[sdot] at (7.80,6) {};
\node[sdot] at (1.00,7) {};
\node[sdot] at (2.00,8) {};
\node[sdot] at (4.40,9) {};
\node[sdot] at (5.40,10) {};
\node[sdot] at (7.80,11) {};
\node[sdot] at (8.80,12) {};
\node[sdot] at (2.00,13) {};
\node[sdot] at (0.00,14) {};
\node[sdot] at (5.40,15) {};
\node[sdot] at (3.40,16) {};
\node[sdot] at (8.80,17) {};
\node[sdot] at (6.80,18) {};
\path let \p1=(current bounding box.east), \p2=(current bounding box.west) in (7.920cm-\x1,\y1) (7.920cm-\x2,\y2);
\end{tikzpicture}

%% file: sections/03_preliminaries.tex
\section{Preliminaries}
\label{sec:prelim}

This section fixes the geometric view of binary search trees and defines the funnel bound together with the count of empty increasing pairs that controls it.

\subsection{Searches as points}
\label{sec:prelim-points}

Keys come from a finite totally ordered set $K$. A \emph{word} over $K$ is a sequence $X=x_1x_2\cdots x_m$ of keys, the searches; keys may repeat unless we say otherwise, and $\abs{X}\coloneqq m$. A word is a permutation if its keys are distinct. For elements $a\le b$ of an ordered set, such as $K$, we write $[a,b]$ for the set of the elements $u$ with $a\le u\le b$, and $(a,b)$, $[a,b)$ and $(a,b]$ for the intervals without one or both ends. The positions of a word $X=x_1\cdots x_m$ written with subscripts are counted from $1$; when we index a word with square brackets, as in $V_\kappa[i]$ (\cref{sec:shuffles}) or $P[i]$ (\cref{sec:levels}), positions are counted from $0$, so $X[0]=x_1$. We place the search $x_t$ at the point $(x_t,t)$ of the plane, keys horizontal and time increasing upwards, and write
\begin{align*}
\Pts(X)\coloneqq\{(x_t,t):1\le t\le m\}.
\end{align*}
For two points $p,q$ let $\rect(p,q)$ be the closed axis-parallel rectangle with corners $p$ and $q$. We say that $q$ lies \emph{to the lower right} of $p$ if $q$ has a larger key and an earlier time than $p$, and similarly for the other three directions.

\begin{definition}[\cite{dhi+09}]
\label{def:satisfied}
A set $S$ of points is \emph{arborally satisfied} if every two points $p,q\in S$ that lie neither in a common row nor in a common column satisfy $\rect(p,q)\cap S\ne\{p,q\}$. For a word $X$ of length $m$, $\OPT(X)$ is the least size of an arborally satisfied set $S\supseteq\Pts(X)$ whose points have times in $\{1,\dots,m\}$.
\end{definition}

In the model of binary search trees of \cite{dhi+09} described in \cref{sec:intro}, the nodes touched by any execution form an arborally satisfied superset of $\Pts(X)$ \cite[Lemma~2.1]{dhi+09}, and every arborally satisfied superset is the set of touched nodes of some execution \cite[Lemma~2.2]{dhi+09}, so $\OPT(X)$ is exactly the least cost of serving $X$, offline and with the best initial tree \cite[Section~2.3]{dhi+09}. Our lower bounds on $\OPT$ are therefore lower bounds on the cost of every binary search tree.

\subsection{The funnel bound}
\label{sec:prelim-funnel}

The definitions in this subsection are those of Lecomte and Weinstein \cite[Definitions~6--8, 13 and 14]{lw20}. For a point $p\in\Pts(X)$, its \emph{left funnel} $F_L(p)$ and \emph{right funnel} $F_R(p)$ are
\begin{align*}
F_L(p)&\coloneqq\{q\in\Pts(X): q\text{ lies to the lower left of }p,\ \rect(p,q)\cap\Pts(X)=\{p,q\}\},\\
F_R(p)&\coloneqq\{q\in\Pts(X): q\text{ lies to the lower right of }p,\ \rect(p,q)\cap\Pts(X)=\{p,q\}\}.
\end{align*}
List the points of $F_L(p)\cup F_R(p)$ in order of time and write $L$ for a point of the left funnel and $R$ for a point of the right funnel. The number $f(p)$ is the number of maximal blocks of equal letters in the resulting word over $\{L,R\}$ (zero for the empty word), and
\begin{align*}
\Funnel(X)\coloneqq\sum_{p\in\Pts(X)}f(p).
\end{align*}
Lecomte and Weinstein note that their definitions of Wilber's two bounds may differ from Wilber's by a constant factor and an additive $O(m)$ \cite[footnote~6]{lw20}.

A word over $\{L,R\}$ with $\ell$ letters $L$ has at most $\ell$ blocks of $L$ and at most $\ell+1$ blocks of $R$, so $f(p)\le 2\abs{F_L(p)}+1$, and only the left funnels need to be counted. An \emph{empty increasing pair} of $X$ is a pair $(s,t)$ of positions such that
\begin{align}
s<t,\qquad x_s<x_t,\qquad x_r\notin[x_s,x_t]\quad(s<r<t),
\label{eq:eip}
\end{align}
that is, such that $(x_s,s)\in F_L((x_t,t))$. Thus
\begin{align*}
\Lambda(X)\coloneqq\sum_{p\in\Pts(X)}\abs{F_L(p)}
\end{align*}
is the number of empty increasing pairs of $X$.

\begin{lemma}
\label{lem:funnel-lambda}
Every word $X$ of length $m$ satisfies $\Funnel(X)\le m+2\Lambda(X)$.
\end{lemma}

\begin{proof}
Sum $f(p)\le2\abs{F_L(p)}+1$ over the $m$ points of $\Pts(X)$.
\end{proof}

%% file: sections/04_rectangles.tex
\section{Independent rectangles}
\label{sec:rect}

This section prepares the lower bound of \cref{thm:main-formal}. The bound is a set of empty rectangles of $\pi_k$, each spanned by two searches of which the earlier one has the larger key, that are pairwise independent in the sense of Demaine et al.\ \cite{dhi+09}. By their lower bound, every such rectangle costs one unit beyond the $m$ searches (\cref{lem:dhikp}). \Cref{lem:staggered} reduces independence to a single configuration of two rectangles, which is all that the later sections check, and \cref{lem:split} carries an independent set from a word with repeated keys to its split, which is a permutation.

\subsection{Empty decreasing pairs and independence}
\label{sec:rect-def}

Let $X\coloneqq x_1\cdots x_m$ be a word. An \emph{empty decreasing pair} of $X$ is a pair $(s,t)$ of positions such that
\begin{align}
s<t,\qquad x_s>x_t,\qquad x_r\notin[x_t,x_s]\quad(s<r<t).
\label{eq:edp}
\end{align}
Its \emph{rectangle} is $B(s,t)\coloneqq[x_t,x_s]\times[s,t]$. The \emph{lower point} $(x_s,s)$ is its lower right corner and the \emph{upper point} $(x_t,t)$ its upper left corner, and by \eqref{eq:edp} it contains no other point of $X$. Empty decreasing pairs are the mirror images of the empty increasing pairs of \cref{sec:prelim-funnel}: $(s,t)$ is one exactly when $(x_s,s)\in F_R((x_t,t))$. For a set $F$ of pairs of positions,
\begin{align*}
\tops(F)\coloneqq\{t:(s,t)\in F\}
\end{align*}
is the set of the positions of its upper points. The same definitions apply to a word whose positions are counted from $0$.

\begin{definition}[{\cite[Section~4.1]{dhi+09}}]
\label{def:independent}
Two empty decreasing pairs $(s,t)$ and $(s',t')$ of $X$ are \emph{independent} if no corner of $B(s,t)$ lies in the interior $(x_{t'},x_{s'})\times(s',t')$ of $B(s',t')$, and symmetrically no corner of $B(s',t')$ lies in the interior of $B(s,t)$. A set of empty decreasing pairs is independent if its pairs are pairwise independent.
\end{definition}

The rectangles of independent pairs may cross, and they may share corners and parts of their boundary. Demaine et al.\ define independence for empty rectangles of both orientations, and their lower bound treats the two orientations separately. We state it for ours. A set $S$ of points is \emph{right-signed satisfied} if
\begin{align*}
\rect((z,t),(y,s))\cap S\ne\{(z,t),(y,s)\}\qquad\text{for all }(z,t),(y,s)\in S\text{ with }z<y,\ s<t,
\end{align*}
that is, if every two of its points of which the lower one lies to the right span a rectangle that contains a third point. Arborally satisfied sets (\cref{def:satisfied}) are right-signed satisfied.

\begin{lemma}[{\cite[Lemma~4.5]{dhi+09}}]
\label{lem:dhikp}
Let $X$ be a permutation of $m$ keys, $F$ an independent set of empty decreasing pairs of $X$, and $S\supseteq\Pts(X)$ a right-signed satisfied set of points. Then $\abs{S}\ge m+\abs{F}$. In particular,
\begin{align*}
\OPT(X)\ge m+\abs{F}.
\end{align*}
\end{lemma}

Demaine et al.\ assume throughout their Section~4 that the access sequence is a permutation (so no two searches share a key), which is why we state \cref{lem:dhikp} for permutations and apply it after splitting repeated keys (\cref{sec:rect-split}). Their bound $\OPT(X)\ge\IRB(X)/2+m$ \cite[Theorem~4.1]{dhi+09} follows from \cref{lem:dhikp}, since one of the two orientations contains at least half of the rectangles of any independent set; when all rectangles already share one orientation, no factor $1/2$ is lost. Their proof marks, for every rectangle of $F$, two horizontally adjacent points of $S$ inside it (using distinctness of the search keys to pick a markable pair) and uses independence to show that no pair is marked twice. Because a row with $\ell$ points of $S$ has $\ell-1$ adjacent pairs and exactly one search, it contains at least as many points of $S\setminus\Pts(X)$ as marked pairs, so $\abs{S\setminus\Pts(X)}\ge\abs F$.

The next lemma is the form in which we check independence: two empty decreasing pairs can fail to be independent in one way only, with one rectangle shifted up and to the right of the other.

\begin{lemma}
\label{lem:staggered}
Two empty decreasing pairs $(s,t)$ and $(s',t')$ of a word $X$ are independent if and only if neither they nor the same two pairs with their roles exchanged satisfy
\begin{align}
x_t<x_{t'}<x_s<x_{s'}\qquad\text{and}\qquad s<s'<t<t'.
\label{eq:staggered}
\end{align}
\end{lemma}

\begin{proof}
If \eqref{eq:staggered} holds, the lower left corner $(x_{t'},s')$ of $B(s',t')$ lies in the interior of $B(s,t)$, so the pairs are not independent. For the converse, suppose that a corner of $B(s',t')$ lies in the interior of $B(s,t)$. The lower and the upper point of $B(s',t')$ are points of $X$, so they do not lie in the interior of $B(s,t)$, which contains no point of $X$ besides its own two. So the corner is the lower left or the upper right one. If it is the lower left corner $(x_{t'},s')$, then
\begin{align*}
x_t<x_{t'}<x_s,\qquad s<s'<t.
\end{align*}
The upper point $(x_{t'},t')$ has its key strictly inside $[x_t,x_s]$ and does not lie in $B(s,t)$, so $t'\notin[s,t]$, and the lower point $(x_{s'},s')$ has its time strictly inside $[s,t]$, so $x_{s'}\notin[x_t,x_s]$. Since $t'>s'>s$ and $x_{s'}>x_{t'}>x_t$,
\begin{align*}
t'>t,\qquad x_{s'}>x_s,
\end{align*}
which is \eqref{eq:staggered}. If it is the upper right corner $(x_{s'},t')$, then $x_t<x_{s'}<x_s$ and $s<t'<t$. The same two observations, now with $s'<t'<t$ and $x_{t'}<x_{s'}<x_s$, give
\begin{align*}
s'<s,\qquad x_{t'}<x_t,
\end{align*}
which is \eqref{eq:staggered} with the roles of the two pairs exchanged. A corner of $B(s,t)$ in the interior of $B(s',t')$ leads to the same two cases by symmetry.
\end{proof}

\subsection{Repeated keys}
\label{sec:rect-split}

Our construction produces a word with repeated keys, and \cref{lem:dhikp} needs a permutation. The \emph{split} of a word $X$ is the permutation $X'$ obtained by giving the $o$-th search of each key $x$ the new key $(x,o)$,
\begin{align*}
x'_t\coloneqq(x_t,o_t),\qquad o_t\coloneqq\abs{\{s\le t:x_s=x_t\}},
\end{align*}
with the new keys ordered by $x$ first and by $o$ second; this is the sequence $\mathrm{split}(X)$ of \cite[Appendix~E]{cgk+15}. The next lemma says that the split keeps both of our counts: the empty increasing pairs grow by less than $m$, and an independent set stays independent.

\begin{lemma}
\label{lem:split}
Let $X$ be a word of length $m$ on $n$ distinct keys and $X'$ its split. Then
\begin{align*}
\Lambda(X')=\Lambda(X)+m-n,
\end{align*}
and every independent set of empty decreasing pairs of $X$ is an independent set of empty decreasing pairs of $X'$.
\end{lemma}

\begin{proof}
For the identity, let $p$ be a point $((x,o),t)$ of $X'$, write $F'_L$ and $F_L$ for left funnels in $X'$ and in $X$, and when $o\ge2$ let $s_0<t$ be the time of the $(o-1)$-th search of $x$. We show
\begin{align*}
F'_L(p)=C_p\cup\{((y,o'),s):(y,s)\in F_L((x,t))\},
\end{align*}
where $C_p\coloneqq\{((x,o-1),s_0)\}$ if $o\ge2$ and $C_p\coloneqq\emptyset$ if $o=1$. An earlier copy $((x,o'),s')$ with $o'<o$ lies in the left funnel of $p$ if and only if $o'=o-1$, since the copies in between lie in the rectangle and no other key lies between $(x,o-1)$ and $(x,o)$. A point $q=((y,o'),s)$ with $y<x$ lies in the left funnel of $p$ in $X'$ if and only if $(y,s)$ lies in the left funnel of $(x,t)$ in $X$. Indeed, with $\varphi$ the map that drops the copy index, $\varphi((u,o''),r)\coloneqq(u,r)$, the order of the keys gives
\begin{align*}
\varphi\bigl(\rect(q,p)\cap\Pts(X')\setminus\{q,p\}\bigr)=\rect((y,s),(x,t))\cap\Pts(X)\setminus\{(y,s),(x,t)\},
\end{align*}
and $\varphi$ is one to one on $\Pts(X')$. The points on the left are the later copies of $y$, the copies of keys strictly between $y$ and $x$, and the earlier copies of $x$, each at a time in $[s,t]$, and these are exactly the points on the right. Pairs of equal keys are not increasing in $X$. So the empty increasing pairs of $X'$ are those of $X$ and one pair for each of the $m-n$ searches that are not the first search of their key.

For the second statement, let $(s,t)$ be an empty decreasing pair of $X$. In $X'$,
\begin{align*}
x'_s>x'_t,\qquad x'_r\notin[x'_t,x'_s]\quad(s<r<t),
\end{align*}
the first because $x_s>x_t$, and the second because $x'_r\in[x'_t,x'_s]$ would give $x_r\in[x_t,x_s]$, against \eqref{eq:edp}. So $(s,t)$ is an empty decreasing pair of $X'$. Now let $(s,t)$ and $(s',t')$ be independent in $X$, and suppose that they are not independent in $X'$. By \cref{lem:staggered} we may assume that \eqref{eq:staggered} holds in $X'$. Dropping the copy indices,
\begin{align*}
x_t\le x_{t'}\le x_s\le x_{s'}\qquad\text{and}\qquad s<s'<t<t'.
\end{align*}
The three inequalities between keys are not all strict, since then \eqref{eq:staggered} would hold in $X$, against \cref{lem:staggered} for the word $X$. We rule out each equality.
\begin{itemize}
\item {\bf Case $x_t=x_{t'}$.} Then $x_t\in[x_{t'},x_{s'}]$ and $s'<t<t'$, against \eqref{eq:edp} for $(s',t')$.
\item {\bf Case $x_s=x_{s'}$.} Then $x_{s'}\in[x_t,x_s]$ and $s<s'<t$, against \eqref{eq:edp} for $(s,t)$.
\item {\bf Case $x_{t'}=x_s$.} Of two searches of the same key, the earlier one has the smaller key in $X'$, and $s<t'$, so $x'_s<x'_{t'}$, against $x'_{t'}<x'_s$ in \eqref{eq:staggered}.
\end{itemize}
So the two pairs are independent in $X'$.
\end{proof}

%% file: sections/05_pair_shuffles.tex
\section{Pair shuffles}
\label{sec:shuffles}

This section defines the composition of words on which the construction is built and proves its two properties. The empty increasing pairs of a pair shuffle are those of its pieces up to an error at descending visits (\cref{prop:pairs}), and the error is at most twice the number of descending visits times the largest number of empty increasing pairs at one point \eqref{eq:err-degree}. Independent sets of empty decreasing pairs of the pieces combine to an independent set of the pair shuffle, with one more pair for every visit that does not descend and is the upper point of a pair of the strip word (\cref{lem:compose,lem:compose-indep}).

\subsection{Definition}
\label{sec:shuffle-def}

Let $\Sigma$ be a finite ordered set, whose elements we call strips, and $U\coloneqq u_1\cdots u_{\abs U}$ a word over $\Sigma$ in which consecutive letters differ. Position $v$ of $U$ is the $v$-th \emph{visit}, and for a strip $\kappa$ let $\occ(\kappa)$ be the number of visits to $\kappa$. For every strip $\kappa$, let $V_\kappa\coloneqq V_\kappa[0]V_\kappa[1]\cdots V_\kappa[\occ(\kappa)]$ be a word over a finite ordered set of labels, of length $\occ(\kappa)+1$, in which consecutive letters differ.

\begin{definition}
\label{def:shuffle}
The \emph{pair shuffle} $Y$ of $U$ and $(V_\kappa)_\kappa$ is the word of length $2\abs U$ over the keys $\Sigma\times\text{labels}$, ordered by strip first and by label second, in which the visit $v$, if it is the $i$-th visit to $\kappa=u_v$, searches two keys:
\begin{align*}
y_{2v-1}\coloneqq(\kappa,V_\kappa[i-1]),\qquad y_{2v}\coloneqq(\kappa,V_\kappa[i]).
\end{align*}
We call $(y_{2v-1},2v-1)$ and $(y_{2v},2v)$ the first point and the second point of the visit. The visit \emph{descends} if $V_\kappa[i]<V_\kappa[i-1]$.
\end{definition}

For a word $W=w_1\cdots w_l$, let $\desc(W)$ be the number of its \emph{descents}, the positions $t<l$ with $w_t>w_{t+1}$. The $i$-th visit to $\kappa$ descends exactly when position $i-1$ is a descent of $V_\kappa$, so the number of descending visits of $Y$ is $\sum_\kappa\desc(V_\kappa)$.

The keys of a strip $\kappa$ form an interval $K_\kappa\coloneqq\{\kappa\}\times\text{labels}$ of the keys of $Y$. Two facts follow at once from the definition. Consecutive letters of $Y$ differ, because the two searches of a visit have different labels and consecutive visits go to different strips. The restriction of $Y$ to $K_\kappa$ is the word of labels
\begin{align}
V_\kappa[0],\ V_\kappa[1],\ V_\kappa[1],\ V_\kappa[2],\ V_\kappa[2],\ \dots,\ V_\kappa[\occ(\kappa)-1],\ V_\kappa[\occ(\kappa)-1],\ V_\kappa[\occ(\kappa)],
\label{eq:doubled}
\end{align}
which we call the \emph{doubled word} of $\kappa$. Merging its consecutive equal letters gives $V_\kappa$, the \emph{in-strip word} of $\kappa$, and $U$ is the \emph{strip word} of $Y$. In particular, the second point of the $i$-th visit to $\kappa$ and the first point of the $(i+1)$-th visit search the same key $(\kappa,V_\kappa[i])$. This is what we mean by saying that consecutive visits to a strip overlap. For a strip $\kappa$ and $1\le i\le\occ(\kappa)$, let $z_\kappa(i)$ be the $i$-th visit to $\kappa$, which searches at the odd time $2z_\kappa(i)-1$ and the even time $2z_\kappa(i)$. In the doubled word \eqref{eq:doubled}, position $i$ of $V_\kappa$ has an earlier copy at the even time $t^-_\kappa(i)$ if $i\ge1$ (the second point of visit $z_\kappa(i)$) and a later copy at the odd time $t^+_\kappa(i)$ if $i<\occ(\kappa)$ (the first point of visit $z_\kappa(i+1)$), where
\begin{align*}
t^-_\kappa(i)\coloneqq2z_\kappa(i),\qquad t^+_\kappa(i)\coloneqq2z_\kappa(i+1)-1.
\end{align*}
The lengths of the pieces add up to the length of $Y$ up to the number of strips:
\begin{align}
\abs{Y}=2\abs{U},\qquad\sum_\kappa\abs{V_\kappa}=\abs{U}+\abs{\Sigma}.
\label{eq:lengths}
\end{align}

\subsection{Empty increasing pairs}
\label{sec:shuffle-pairs}

An empty increasing pair of $Y$ joins two points inside a strip or two points in different strips. The next lemma characterizes emptiness across two strips.

\begin{lemma}
\label{lem:cross-empty}
Let $q$ be a point of $Y$ in the strip $\kappa_1$ at the visit $v_1$, and $p$ a point in a strip $\kappa_2>\kappa_1$ at a visit $v_2>v_1$. Then $\rect(q,p)$ contains no other point of $Y$ if and only if the following three conditions hold.
\begin{itemize}
\item {\bf No strip in between.} No visit $w$ with $v_1<w<v_2$ goes to a strip strictly between $\kappa_1$ and $\kappa_2$.
\item {\bf Smaller in the lower strip.} Every point of the strip $\kappa_1$ strictly between $q$ and $p$ in time has a smaller label than $q$.
\item {\bf Larger in the upper strip.} Every point of the strip $\kappa_2$ strictly between $q$ and $p$ in time has a larger label than $p$.
\end{itemize}
\end{lemma}

\begin{proof}
Keys are ordered by strip first, so the keys of $\rect(q,p)$ are the keys of the strips strictly between $\kappa_1$ and $\kappa_2$, the keys of $\kappa_1$ with a label at least that of $q$, and the keys of $\kappa_2$ with a label at most that of $p$. The points of $Y$ strictly between $q$ and $p$ in time belong to the visits strictly between $v_1$ and $v_2$ and to the visits $v_1$ and $v_2$ themselves, which lie in the strips $\kappa_1$ and $\kappa_2$.
\end{proof}

Let $\err(Y)$ be the number of empty increasing pairs $(s,t)$ of $Y$ whose points lie in different strips and such that $s$ is the first time of a descending visit or $t$ is the second time of a descending visit.

\begin{proposition}
\label{prop:pairs}
Every pair shuffle $Y$ of $U$ and $(V_\kappa)_\kappa$ satisfies
\begin{align*}
\Lambda(Y)=\Lambda(U)+\sum_\kappa\Lambda(V_\kappa)+\err(Y).
\end{align*}
\end{proposition}

\begin{proof}
We count the empty increasing pairs $(s,t)$ of $Y$ with both points in one strip and those with points in two strips separately.

{\bf Pairs inside a strip.} Fix a strip $\kappa$, let $o\coloneqq\occ(\kappa)$, and write $t^-\coloneqq t^-_\kappa$ and $t^+\coloneqq t^+_\kappa$ for the copy times (\cref{sec:shuffle-def}). A rectangle with both corners in $K_\kappa$ contains only keys of $K_\kappa$, so these pairs are the empty increasing pairs of the doubled word \eqref{eq:doubled}. The $i$-th visit to $\kappa$ searches at the times $t^+(i-1)$ and $t^-(i)$, so in order of time the points of $K_\kappa$ lie at
\begin{align*}
t^+(0)<t^-(1)<t^+(1)<t^-(2)<\cdots<t^+(o-1)<t^-(o).
\end{align*}
We show that the empty increasing pairs inside $K_\kappa$ are exactly
\begin{align}
\bigl\{\bigl(t^+(i'),t^-(i)\bigr):(i',i)\text{ is an empty increasing pair of }V_\kappa\bigr\}.
\label{eq:pairs-inside}
\end{align}
An increasing pair inside $K_\kappa$ joins a copy of a position $i'$ with a copy of a later position $i>i'$, with $V_\kappa[i']<V_\kappa[i]$. All copies of $i'$ precede all copies of $i$. If the earlier time is $t^-(i')$, then the later copy $t^+(i')$, which exists since $i'<i\le o$, has the same key and lies strictly between the two times, so it lies in the rectangle. Symmetrically, if the later time is $t^+(i)$, then the earlier copy $t^-(i)$ lies in the rectangle. Thus any empty pair has the form $(t^+(i'),t^-(i))$. The points of $K_\kappa$ strictly between $t^+(i')$ and $t^-(i)$ in time are the copies of the positions $l$ with $i'<l<i$, so $(t^+(i'),t^-(i))$ satisfies \eqref{eq:eip} in $Y$ if and only if
\begin{align*}
V_\kappa[l]\notin\bigl[V_\kappa[i'],V_\kappa[i]\bigr]\qquad(i'<l<i),
\end{align*}
that is, if and only if $(i',i)$ is an empty increasing pair of $V_\kappa$. This proves \eqref{eq:pairs-inside}, so strip $\kappa$ contributes $\Lambda(V_\kappa)$.

{\bf Pairs across strips.} Let $(s,t)$ be an empty increasing pair of $Y$ across strips that is not counted by $\err(Y)$, with $q\coloneqq(y_s,s)$ in strip $\kappa_1$ at visit $v_1$ and $p\coloneqq(y_t,t)$ in strip $\kappa_2$ at visit $v_2$. Keys are ordered by strip first and $y_s<y_t$, so $\kappa_1<\kappa_2$, and $v_1<v_2$ because $s<t$ and the two points lie in different visits. We show
\begin{align}
s=2v_1,\qquad t=2v_2-1,\qquad u_w\notin[\kappa_1,\kappa_2]\quad(v_1<w<v_2),
\label{eq:pairs-across}
\end{align}
where the third relation says that $(v_1,v_2)$ is an empty increasing pair of $U$. We apply the three conditions of \cref{lem:cross-empty} to $q$ and $p$.

If $t=2v_2$, then $v_2$ does not descend, since $(s,t)$ is not counted by $\err(Y)$. Because consecutive letters of $V_{\kappa_2}$ differ, the first point $(y_{2v_2-1},2v_2-1)$ of $v_2$ has a smaller label than $p$ and lies strictly between $s$ and $t$ in time, contradicting \emph{larger in the upper strip}. Hence $t=2v_2-1$. Symmetrically, \emph{smaller in the lower strip} forces $s=2v_1$.

For the third relation, let $v^-$ be the last visit to $\kappa_2$ before $v_2$, if there is one. By the overlap, its second point searches the same key as $p$:
\begin{align*}
y_{2v^-}=y_{2v_2-1}.
\end{align*}
If $v_1<v^-$, then time $2v^-$ lies strictly between $s=2v_1$ and $t=2v_2-1$ and carries the label of $p$, contradicting \emph{larger in the upper strip}; thus no visit strictly between $v_1$ and $v_2$ goes to $\kappa_2$. Symmetrically, considering the first visit to $\kappa_1$ after $v_1$, no visit strictly between $v_1$ and $v_2$ goes to $\kappa_1$. Combined with \emph{no strip in between}, this establishes \eqref{eq:pairs-across}.

Conversely, let $(v_1,v_2)$ be an empty increasing pair of $U$. Every point strictly between the times $2v_1$ and $2v_2-1$ belongs to a visit $w$ with $v_1<w<v_2$, whose strip lies outside $[u_{v_1},u_{v_2}]$, so $(2v_1,2v_2-1)$ is an empty increasing pair of $Y$. It is not counted by $\err(Y)$, because $2v_1$ is a second time and $2v_2-1$ is a first time. Thus $(v_1,v_2)\mapsto(2v_1,2v_2-1)$ is a bijection between the empty increasing pairs of $U$ and the cross-strip pairs not counted by $\err(Y)$. Together with \eqref{eq:pairs-inside}, this proves the proposition.
\end{proof}

To bound the error, let $A(Y)$ be the largest number of empty increasing pairs of $Y$ that share a common later position, or share a common earlier position. Every pair counted by $\err(Y)$ either starts at the first time of a descending visit or ends at the second time of one, so
\begin{align}
\err(Y)\le2A(Y)\cdot\sum_\kappa\desc(V_\kappa).
\label{eq:err-degree}
\end{align}

\subsection{Independent sets of empty decreasing pairs}
\label{sec:shuffle-rect}

We now turn to the lower bound. Given independent sets of empty decreasing pairs of the strip word and of the in-strip words (\cref{def:independent}), the next two lemmas build one of the pair shuffle. It contains a copy of every given pair, and it uses one pair of the strip word a second time at every visit that does not descend. This second use is the gain of a pair shuffle, and it is the only place where the overlap of consecutive visits enters the lower bound. At a descending visit, the two points of the visit form an empty decreasing pair, which we add as well, so that both points of every visit that is an upper point of the strip word become upper points.

We use the notation $z_\kappa(i)$, $t^-_\kappa$ and $t^+_\kappa$ of \cref{sec:shuffle-def}, with the positions of $V_\kappa$ counted from $0$ as in \cref{def:shuffle}. The set $F$ of the next lemma has four kinds of pairs, listed in \eqref{eq:compose-sets}. A pair $(i',i)$ of an in-strip word $V_\kappa$ is copied to the later copy of position $i'$ and the earlier copy of position $i$, namely $(t^+_\kappa(i'),t^-_\kappa(i))=(2z_\kappa(i'+1)-1,2z_\kappa(i))$; both visits exist because $0\le i'<i\le\occ(\kappa)$. A pair $(w,v)$ of the strip word is copied to $(2w,2v-1)$, joining the second point of visit $w$ with the first point of visit $v$. At a non-descending visit $v$, the second use $(2w_v,2v)$ joins the second point of $v$ with the second point of the earliest visit $w_v$ such that $(w_v,v)\in F_U$. At a descending visit $v$, the pair $(2v-1,2v)$ of its two points is added (it may already belong to the in-strip copies).

\begin{lemma}
\label{lem:compose}
Let $Y$ be the pair shuffle of $U$ and $(V_\kappa)_\kappa$. Let $F_U$ be an independent set of empty decreasing pairs of $U$ and, for every strip $\kappa$, let $F_\kappa$ be an independent set of empty decreasing pairs of $V_\kappa$. Let $T\coloneqq\tops(F_U)$, for $v\in T$ let $w_v\coloneqq\min\{w:(w,v)\in F_U\}$, let $T^{\uparrow}$ be the set of the visits in $T$ that do not descend, and let $\Delta$ be the set of all descending visits. Then the first three of the sets
\begin{align}
F^{\mathrm{in}}&\coloneqq\bigl\{\bigl(t^+_\kappa(i'),\ t^-_\kappa(i)\bigr):\kappa\in\Sigma,\ (i',i)\in F_\kappa\bigr\},\notag\\
F^{\mathrm{st}}&\coloneqq\bigl\{(2w,\ 2v-1):(w,v)\in F_U\bigr\},\notag\\
F^{+}&\coloneqq\bigl\{(2w_v,\ 2v):v\in T^{\uparrow}\bigr\},\notag\\
F^{-}&\coloneqq\bigl\{(2v-1,\ 2v):v\in\Delta\bigr\}
\label{eq:compose-sets}
\end{align}
are disjoint. Their union $F$ consists of empty decreasing pairs of $Y$, and
\begin{align}
\abs{F}\ge\abs{F_U}+\sum_\kappa\abs{F_\kappa}+\abs{T^{\uparrow}},\qquad\tops(F)\supseteq\{2v-1:v\in T\}\cup\{2v:v\in T\}.
\label{eq:compose-count}
\end{align}
In particular,
\begin{align}
\abs{F}&\ge\abs{F_U}+\sum_\kappa\abs{F_\kappa}+\abs{T}-\sum_\kappa\desc(V_\kappa),\notag\\
\abs{\tops(F)}&\ge2\abs{T}.
\label{eq:compose-bounds}
\end{align}
\end{lemma}

\begin{proof}
We first count. Each of the first three maps in \eqref{eq:compose-sets} is one to one, and the three sets are disjoint. Indeed, the upper points of $F^{\mathrm{st}}$ are first points of visits, at odd times, and those of the other two sets are second points, at even times:
\begin{align*}
\tops(F^{\mathrm{st}})\subseteq\{2v-1:v\le\abs U\},\qquad\tops(F^{\mathrm{in}})\cup\tops(F^+)\subseteq\{2v:v\le\abs U\}.
\end{align*}
A pair of $F^{\mathrm{in}}$ has both points in one strip, and a pair of $F^+$ has its points in two strips. This gives the count in \eqref{eq:compose-count}. Both points of every visit $v\in T$ are upper points:
\begin{align*}
2v-1\in\tops(F^{\mathrm{st}}),\qquad2v\in\begin{cases}\tops(F^+)&(v\text{ does not descend}),\\\tops(F^-)&(v\text{ descends}),\end{cases}
\end{align*}
which is the inclusion in \eqref{eq:compose-count}. The bounds \eqref{eq:compose-bounds} follow from $\abs{T^{\uparrow}}\ge\abs T-\abs\Delta$ and $\abs\Delta=\sum_\kappa\desc(V_\kappa)$.

Next we show that the pairs of $F$ are empty decreasing pairs of $Y$. For a pair of $F^{\mathrm{in}}$ in the strip $\kappa$, with $(i',i)\in F_\kappa$, both points lie in $K_\kappa$ and have the labels $V_\kappa[i']>V_\kappa[i]$. By \eqref{eq:doubled}, the points of $K_\kappa$ strictly between them in time are the copies of the positions $j$ with $i'<j<i$, and by \eqref{eq:edp} for $(i',i)$ in $V_\kappa$,
\begin{align*}
V_\kappa[j]\notin\bigl[V_\kappa[i],V_\kappa[i']\bigr]\qquad(i'<j<i).
\end{align*}
The points of other strips have keys outside $K_\kappa$, so \eqref{eq:edp} holds in $Y$. For a pair $(w,v)\in F_U$, \eqref{eq:edp} in $U$ reads
\begin{align}
w<v,\qquad u_w>u_v,\qquad u_r\notin[u_v,u_w]\quad(w<r<v).
\label{eq:strip-empty}
\end{align}
Let $K_{[u_v,u_w]}$ be the set of the keys in the strips of $[u_v,u_w]$. Keys are ordered by strip first, so the key ranges of the two pairs that $(w,v)$ gives lie in this set, while by \eqref{eq:strip-empty} the visits strictly between $w$ and $v$ avoid it:
\begin{align*}
[y_{2v-1},y_{2w}]\cup[y_{2v},y_{2w}]\subseteq K_{[u_v,u_w]},\qquad y_{2r-1},\ y_{2r}\notin K_{[u_v,u_w]}\quad(w<r<v).
\end{align*}
The points strictly between the times $2w$ and $2v-1$ belong to these visits, so $(2w,2v-1)$ is an empty decreasing pair. Between the times $2w$ and $2v$ lies one more point, the first point of $v$. If $v$ does not descend, then $y_{2v-1}<y_{2v}$, so this point lies below $[y_{2v},y_{2w}]$, and $(2w,2v)$ is an empty decreasing pair as well. Finally, a pair of $F^-$ joins two consecutive times, so no point lies between them, and it is decreasing because its visit descends.
\end{proof}

Independence is the main point. A pair of $F$ that comes from a pair of an in-strip word is kept apart from the other pairs by the strips. The real case is two pairs that come from pairs of the strip word, and it is the only place where the choice of $w_v$ and the overlap of consecutive visits are needed.

\begin{lemma}
\label{lem:compose-indep}
The set $F$ of \cref{lem:compose} is independent.
\end{lemma}

\begin{proof}
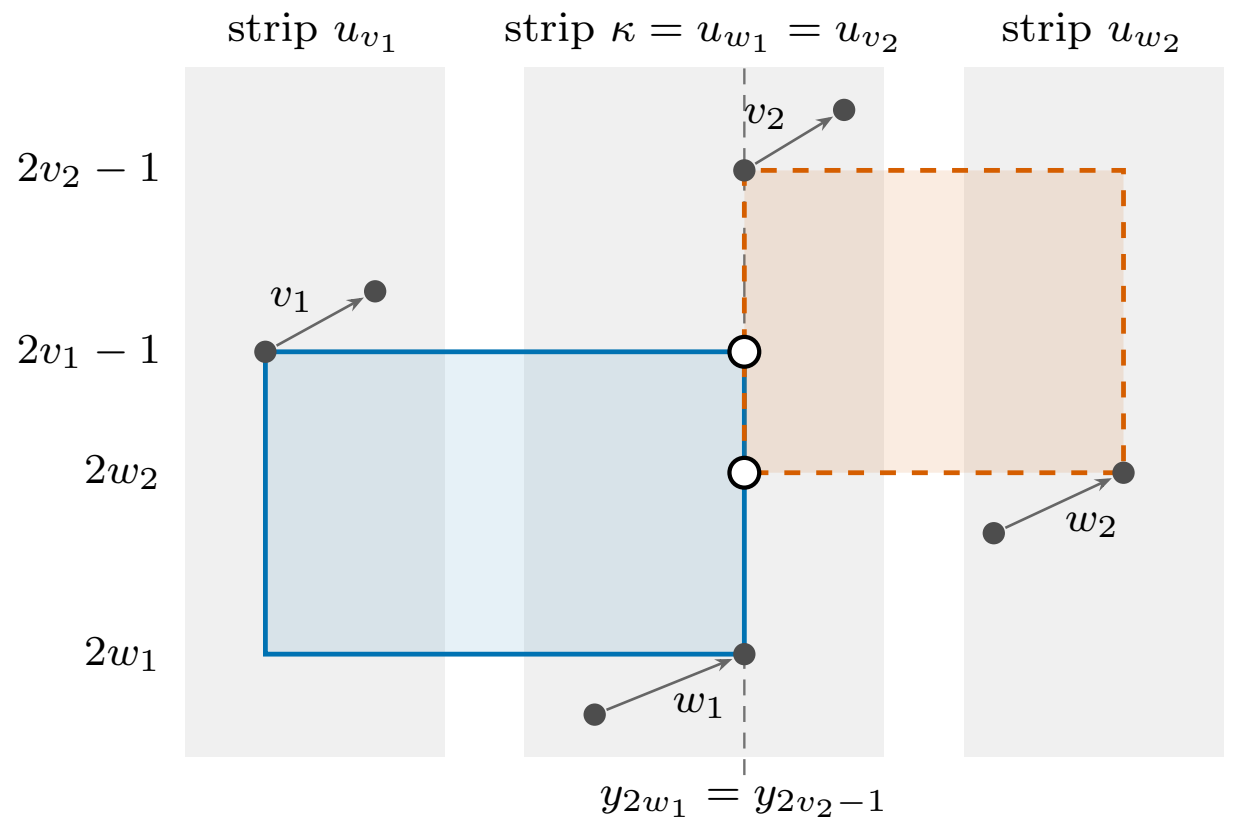
\begin{figure}[t]
\centering
\input{figures/two_strip}
\caption{The last configuration in Case~3 of the proof of \cref{lem:compose-indep}, with keys horizontal and time upwards. The grey bands are strips, and an arrow joins the two points of a visit; the visits between them, which do not go to the strip $\kappa$, are omitted. The rectangles are pairs of $F^{\mathrm{st}}\cup F^+$ that come from $(w_1,v_1)$ (solid, shown for $t=2v_1-1$) and from $(w_2,v_2)$ (dashed, shown for $t'=2v_2-1$), whose strips satisfy $u_{v_1}<u_{v_2}=u_{w_1}<u_{w_2}$. By the overlap, the first point of $v_2$ repeats the key of the second point of $w_1$, so the two rectangles share part of a vertical edge, and the corners marked by circles lie on the boundary of the other rectangle, not in its interior as in \cref{fig:config}(a); if $t'=2v_2$, then $v_2$ ascends to the right of the dashed line and the rectangles do not meet.}
\label{fig:two-strip}
\end{figure}
By \cref{lem:staggered}, it suffices to show that no two pairs $(s,t)$ and $(s',t')$ of $F$ satisfy
\begin{align}
y_t<y_{t'}<y_s<y_{s'}\qquad\text{and}\qquad s<s'<t<t'.
\label{eq:stag-Y}
\end{align}
In \eqref{eq:stag-Y}, each of the two pairs has a time strictly between the two times of the other, $t$ and $s'$, and a key strictly between the two keys of the other, $y_s$ and $y_{t'}$. A pair of $F^-$ joins two consecutive times, so it has no time strictly between its own two times, and both pairs lie in $F^{\mathrm{in}}\cup F^{\mathrm{st}}\cup F^+$. We call the pairs of $F^{\mathrm{st}}\cup F^+$ \emph{strip pairs}. The strip pair of $(w,v)\in F_U$ has its lower point at time $2w$ and its upper point at time $2v-1$ or $2v$, and $(w,v)$ satisfies \eqref{eq:strip-empty}. Suppose that \eqref{eq:stag-Y} holds. There are three cases.

{\bf Case 1.} Both pairs lie in $F^{\mathrm{in}}$. If they belong to different strips $\kappa\ne\kappa'$, then $y_t,y_s\in K_\kappa$ and $y_{t'}\in K_{\kappa'}$, and $y_t<y_{t'}<y_s$ is impossible because $K_\kappa$ is an interval. Let both belong to $\kappa$, with $(i_1',i_1),(i_2',i_2)\in F_\kappa$, and write $t^-\coloneqq t^-_\kappa$ and $t^+\coloneqq t^+_\kappa$, so that $(s,t)=(t^+(i_1'),t^-(i_1))$ and $(s',t')=(t^+(i_2'),t^-(i_2))$. For all positions $j,j'$,
\begin{align*}
t^+(j)<t^+(j')\iff j<j',\qquad t^-(j)<t^-(j')\iff j<j',\qquad t^+(j)<t^-(j')\iff j<j',
\end{align*}
the last because $2z_\kappa(j+1)-1<2z_\kappa(j')$ holds if and only if $j+1\le j'$. So \eqref{eq:stag-Y} reads
\begin{align*}
V_\kappa[i_1]<V_\kappa[i_2]<V_\kappa[i_1']<V_\kappa[i_2']\qquad\text{and}\qquad i_1'<i_2'<i_1<i_2,
\end{align*}
which is \eqref{eq:staggered} for the two pairs in $V_\kappa$, against the independence of $F_\kappa$.

{\bf Case 2.} One pair lies in $F^{\mathrm{in}}$ in the strip $\kappa$, and the other is the strip pair of $(w,v)$. Both keys of the in-strip pair lie in the interval $K_\kappa$, and a key of the strip pair lies between them, so $\kappa$ is the strip $u_w$ or $u_v$ of a point of the strip pair, and $\kappa\in[u_v,u_w]$. A time $\tau$ of the in-strip pair lies strictly between the two times of the strip pair, and it is a time of a visit $x$ to $\kappa$:
\begin{align*}
2w<\tau<2v,\qquad\tau\in\{2x-1,2x\}.
\end{align*}
So $w<x\le v$, and since by \eqref{eq:strip-empty} no visit strictly between $w$ and $v$ goes to $\kappa$, we get $x=v$ and $\tau=2v-1$. Because every pair of $F^{\mathrm{in}}$ starts at an odd time $t^+_\kappa(i')$ and ends at an even time $t^-_\kappa(i)$, this odd time $\tau=2v-1$ is the first time of the in-strip pair. Of the two pairs in \eqref{eq:stag-Y}, only $(s',t')$ has its first time between the times of the other, so the in-strip pair is $(s',t')$ with $s'=2v-1$, and since $s'<t\le2v$, the strip pair is $(s,t)=(2w,2v)$. Then $y_{s'}$ lies in the strip $u_v$ and $y_s$ in the strip $u_w>u_v$, against $y_s<y_{s'}$ in \eqref{eq:stag-Y}.

{\bf Case 3.} Both are strip pairs. Let $(s,t)$ be the strip pair of $(w_1,v_1)$ and $(s',t')$ that of $(w_2,v_2)$. Then $s=2w_1$, $s'=2w_2$, the time $t\in\{2v_1-1,2v_1\}$ belongs to the visit $v_1$ and $t'\in\{2v_2-1,2v_2\}$ to $v_2$, so $s<s'<t<t'$ in \eqref{eq:stag-Y} gives $w_1<w_2<v_1\le v_2$. Keys are ordered by strip first, so $y_t<y_{t'}<y_s<y_{s'}$ gives the same order on their strips, possibly with equalities:
\begin{align}
w_1<w_2<v_1\le v_2,\qquad u_{v_1}\le u_{v_2}\le u_{w_1}\le u_{w_2}.
\label{eq:two-strip}
\end{align}
The visit $w_2$ lies strictly between $w_1$ and $v_1$, so $u_{w_2}\notin[u_{v_1},u_{w_1}]$ by \eqref{eq:strip-empty} for $(w_1,v_1)$, and hence
\begin{align*}
u_{w_1}<u_{w_2}.
\end{align*}
If $v_1=v_2\eqqcolon v$, then $t<t'$ forces $t=2v-1$ and $t'=2v$. So $(s',t')$ is the pair of $F^+$ at $v$, and $w_2=w_v\le w_1$ by the choice of $w_v$, against \eqref{eq:two-strip}. This is where the least position $w_v$ is needed.

Let $v_1<v_2$. Then $v_1$ lies strictly between $w_2$ and $v_2$, so $u_{v_1}\notin[u_{v_2},u_{w_2}]$ by \eqref{eq:strip-empty} for $(w_2,v_2)$, and hence $u_{v_1}<u_{v_2}$. If also $u_{v_2}<u_{w_1}$, then
\begin{align*}
u_{v_1}<u_{v_2}<u_{w_1}<u_{w_2}\qquad\text{and}\qquad w_1<w_2<v_1<v_2,
\end{align*}
which is \eqref{eq:staggered} for $(w_1,v_1)$ and $(w_2,v_2)$ in $U$, against the independence of $F_U$. So $u_{v_2}=u_{w_1}\eqqcolon\kappa$, as in \cref{fig:two-strip}. No visit strictly between $w_1$ and $v_2$ goes to $\kappa$:
\begin{align*}
u_r\notin[u_{v_1},u_{w_1}]\quad(w_1<r<v_1),\qquad u_{v_1}<\kappa,\qquad u_r\notin[u_{v_2},u_{w_2}]\quad(v_1<r<v_2),
\end{align*}
by \eqref{eq:strip-empty} for the two pairs, and $\kappa$ lies in both intervals. So $v_2$ is the visit to $\kappa$ that follows $w_1$, and by the overlap of consecutive visits (\cref{sec:shuffle-def}) its first point repeats the key of the second point of $w_1$. The time $t'$ is $2v_2-1$, or it is $2v_2$ and $v_2$ does not descend, so
\begin{align*}
y_{t'}\ge y_{2v_2-1}=y_{2w_1}=y_s,
\end{align*}
against $y_{t'}<y_s$ in \eqref{eq:stag-Y}. This is where the overlap is needed.
\end{proof}

The choice of $w_v$ cannot be dropped. If $(w_1,v)$ and $(w_2,v)$ lie in $F_U$ with $w_1<w_2$, and $v$ does not descend, then $(2w_2,2v)$ is an empty decreasing pair of $Y$ that satisfies \eqref{eq:stag-Y} together with the pair $(2w_1,2v-1)$ of $F^{\mathrm{st}}$, since $u_{w_1}<u_{w_2}$ by \eqref{eq:strip-empty}.

%% file: figures/two_strip.tex
\begin{tikzpicture}[x=0.66cm,y=0.40cm]
\fill[black!6] (0.0,0.3) rectangle (2.6,11.7);
\fill[black!6] (3.4,0.3) rectangle (7.0,11.7);
\fill[black!6] (7.8,0.3) rectangle (10.4,11.7);
\node[figtxt,anchor=south] at (1.3,11.85) {strip $u_{v_1}$};
\node[figtxt,anchor=south] at (5.2,11.85) {strip $\kappa=u_{w_1}=u_{v_2}$};
\node[figtxt,anchor=south] at (9.1,11.85) {strip $u_{w_2}$};
\draw[black!55,line width=0.45pt,densely dashed] (5.6,0.0) -- (5.6,11.7);
\node[figtxt,anchor=north] at (5.6,-0.08) {$y_{2w_1}=y_{2v_2-1}$};
\node[figtxt,anchor=east] at (-0.18,2) {$2w_1$};
\node[figtxt,anchor=east] at (-0.18,5) {$2w_2$};
\node[figtxt,anchor=east] at (-0.18,7) {$2v_1-1$};
\node[figtxt,anchor=east] at (-0.18,10) {$2v_2-1$};
\draw[rst] (0.8,2) rectangle (5.6,7);
\draw[rplus] (5.6,5) rectangle (9.4,10);
\draw[varrow] (4.1,1) -- (5.6,2);
\draw[varrow] (8.1,4) -- (9.4,5);
\draw[varrow] (0.8,7) -- (1.9,8);
\draw[varrow] (5.6,10) -- (6.6,11);
\foreach \p in {(4.1,1),(5.6,2),(8.1,4),(9.4,5),(0.8,7),(1.9,8),(5.6,10),(6.6,11)} \node[sdot] at \p {};
\node[figtxt,anchor=north west] at (4.82,1.45) {$w_1$};
\node[figtxt,anchor=north west] at (8.75,4.45) {$w_2$};
\node[figtxt,anchor=south east] at (1.35,7.55) {$v_1$};
\node[figtxt,anchor=south east] at (6.10,10.58) {$v_2$};
\node[cdot] at (5.6,5) {};
\node[cdot] at (5.6,7) {};
\end{tikzpicture}

%% file: sections/06_level_words.tex
\section{Level words and the separation}
\label{sec:levels}

This section builds the separating permutations $\pi_k$ by iterating pair shuffles, and proves \cref{thm:main-formal}, the formal version of \cref{thm:main-informal}. The proof needs two bounds per search for the level-$j$ words: at most $1+O(6^j/g)$ empty increasing pairs (\cref{prop:lambda-level}), which is less than $2$ for the base of \cref{thm:main-formal}, and an independent set of at least $j/2$ empty decreasing pairs (\cref{prop:rect-level}). Throughout, for a word $W$ we use the count per search
\begin{align*}
\lambda(W)\coloneqq\frac{\Lambda(W)}{\abs{W}}.
\end{align*}

\subsection{Definition}
\label{sec:level-def}

Level words are pair shuffles nested $j$ times, with cyclic scans at the bottom. A pair shuffle needs an in-strip word with one more letter than its strip has visits, and the definition provides it by repeating the first letter of the label word at its end.

\begin{definition}
\label{def:level}
Let $g\ge2$, the \emph{base}. For $r\ge1$, the level-$0$ word $W_0[r]\coloneqq(0\ 1\ \cdots\ g-1)^r$ is a cyclic scan of the keys $0,\dots,g-1$ with $r$ rounds. For $j\ge1$ and $r\ge1$, the level-$j$ word $W_j[r]$ is the pair shuffle of the strip word and the in-strip words
\begin{align*}
U\coloneqq W_{j-1}\bigl[g^{2^{j-1}}r\bigr],\qquad V_\kappa\coloneqq V'\coloneqq V\ V[0]\quad\text{for every strip }\kappa,
\end{align*}
where $V\coloneqq W_{j-1}[r]$ is the \emph{label word} and $V\ V[0]$ is $V$ followed by its first letter.
\end{definition}

For a word $Z=z_1\cdots z_l$, a \emph{cyclic descent} is a position $t\le l$ with $z_t>z_{t+1}$, where $z_{l+1}\coloneqq z_1$, and $\desc^\circ(Z)$ is the number of cyclic descents. The next lemma collects the facts about level words that the proofs use: their sizes, that they repeat a fixed period, and that exactly a fraction $1/g$ of their cyclic steps descend.

\begin{lemma}
\label{lem:level-basic}
For all $j\ge0$ and $r\ge1$, the word $W\coloneqq W_j[r]$ is defined, its consecutive letters differ, its last letter differs from its first, and its first letter is its smallest key. It has $g^{2^j}$ keys, each of which occurs $2^jr$ times, and
\begin{align}
\abs{W}=2^jg^{2^j}r,\qquad W_j[r]=\bigl(W_j[1]\bigr)^r,\qquad\desc^\circ(W)=\frac{\abs W}g.
\label{eq:level-basic}
\end{align}
For $j\ge1$, with $U$, $V$ and $q\coloneqq g^{2^{j-1}}$ strips as in \cref{def:level}, the word $W$ has
\begin{align}
\abs{U}=q\abs{V},\qquad\abs{W}=2q\abs{V},\qquad\Lambda(V')=\Lambda(V),\qquad\sum_\kappa\desc(V_\kappa)=\frac{\abs W}{2g}.
\label{eq:level-lengths}
\end{align}
\end{lemma}

\begin{proof}
We induct on $j$. The cyclic scan $W_0[r]$ has $g$ keys, each occurring $r$ times. It starts with its smallest key $0$ and ends with $g-1\ne0$, and its cyclic descents are the $r$ steps from $g-1$ to $0$, so $\desc^\circ(W_0[r])=r=\abs{W_0[r]}/g$.

Let $j\ge1$, and write $P\coloneqq W_{j-1}[1]$ and $d\coloneqq\abs P=2^{j-1}g^{2^{j-1}}$. By the induction hypothesis $U=P^{qr}$ and $V=P^r$, and every key of $P$ occurs $2^{j-1}$ times in $P$.

We first show that $W$ is defined. Every strip occurs in $U$
\begin{align*}
q2^{j-1}r=dr=\abs V
\end{align*}
times, so the in-strip word $V'=V\ V[0]$ has the length that \cref{def:shuffle} requires. Its consecutive letters differ, because those of $V$ do and the last letter of $V$ differs from $V[0]$. So $W$ is defined, and its consecutive letters differ (\cref{sec:shuffle-def}). Its first letter is the pair of the first letters of $U$ and $V$, both smallest keys, so it is the smallest key of $W$. Its last letter lies in the last strip of $U$, which differs from the first.

Next we count. In the doubled word \eqref{eq:doubled} of a strip, every position of $V'$ except the two ends appears twice, and both ends carry the label $V[0]$. So a key $(\kappa,h)$ occurs twice as often in $W$ as $h$ in $V$, the keys of $W$ are the $q^2=g^{2^j}$ pairs of a strip and a label, and
\begin{align*}
\mathrm{occ}_W(\kappa,h)=2\cdot\mathrm{occ}_V(h)=2^jr,\qquad\abs W=2\abs U=2qdr=2^jg^{2^j}r,
\end{align*}
where $\mathrm{occ}$ counts occurrences. This is the first relation of \eqref{eq:level-basic}.

For the period, split the visits of $W$ into $r$ blocks of $qd$ consecutive visits. The strip word of each block is $P^q$, in which every strip occurs $q2^{j-1}=d$ times. So the $i$-th visit to a strip in the $\rho$-th block is its $\bigl((\rho-1)d+i\bigr)$-th visit overall, and since $V'=P^r\ P[0]$, its two points have the labels
\begin{align*}
V'\bigl[(\rho-1)d+i-1\bigr]=P[i-1],\qquad V'\bigl[(\rho-1)d+i\bigr]=P[i\bmod d].
\end{align*}
These do not depend on $\rho$, so every block is the same word $W_j[1]$.

For the cyclic descents, we count the cyclic steps of $W$ in two kinds. A step between two consecutive visits, including the step from the last visit to the first, joins two different strips, so it descends exactly when the corresponding cyclic step of $U$ does. A step inside a visit descends exactly when the visit descends, and the visits to one strip descend at the descents of $V'=V\ V[0]$, which are the cyclic descents of $V$. Hence
\begin{align*}
\desc^\circ(W)&=\desc^\circ(U)+q\desc^\circ(V)\\*
&=\frac{\abs U}g+\frac{q\abs V}g\\*
&=\frac{\abs W}g.
\end{align*}
The first equality adds the two kinds of steps, the second is the induction hypothesis, and the last uses $\abs W=2\abs U=2q\abs V$. The steps inside visits alone give the last relation of \eqref{eq:level-lengths}:
\begin{align*}
\sum_\kappa\desc(V_\kappa)=q\desc^\circ(V)=\frac{\abs W}{2g}.
\end{align*}

Finally, the appended letter $V[0]$ is the smallest key of $V'$ and its last letter, so it is the earlier point of no pair and the later point of no increasing pair, and $\Lambda(V')=\Lambda(V)$.
\end{proof}

We write $P_j\coloneqq W_j[1]$ for the period of the level-$j$ words, and $d_j\coloneqq\abs{P_j}=2^jg^{2^j}$. By \eqref{eq:level-basic}, every $d_j$ consecutive positions of $W_j[r]$ contain every key of $P_j$.

\subsection{Empty increasing pairs of level words}
\label{sec:level-pairs}

By \cref{prop:pairs} and \eqref{eq:err-degree}, the empty increasing pairs of a level word are additive up to an error of at most $2A(W)$ per descending visit, where $A(W)$ is the largest number of empty increasing pairs at one position. This subsection proves $A(W_j[r])\le2\cdot6^j$. \Cref{lem:pairs-records} bounds the pairs at a position by the records, defined next, of a short \emph{factor} (contiguous subword) of the word. \Cref{lem:records-strip} bounds the records in a single strip, \cref{lem:records-far} bounds those beyond the nearest strip below or above the key that the factor visits, and \cref{lem:records-recursion} combines the two bounds by induction on $j$.

For a word $Z=z_1\cdots z_l$, the \emph{lower records} of $Z$ below a key $x$ or below $x=+\infty$, and the \emph{upper records} of $Z$ above a key $x$ or above $x=-\infty$, are the sets of positions
\begin{align}
R^-(Z,x)&\coloneqq\bigl\{s:z_s<x\ \text{ and }\ z_{s'}\notin(z_s,x)\ \text{ for all }s'>s\bigr\},\notag\\
R^+(Z,x)&\coloneqq\bigl\{s:z_s>x\ \text{ and }\ z_{s'}\notin(x,z_s)\ \text{ for all }s'<s\bigr\}.
\label{eq:records}
\end{align}
A lower record is a position below $x$ that is not followed by a key strictly between its own key and $x$. A later position with the same key does not disqualify it. For example, for $Z=2\ 1\ 2$ and $x=3$, both searches of $2$ are lower records, and the search of $1$ is not, since it is followed by the key $2\in(1,3)$:
\begin{align*}
R^-(2\ 1\ 2,\ 3)=\{1,3\}.
\end{align*}
For $j\ge0$, let $H_j$ be the largest size of $R^-(Z,x)$ or $R^+(Z,x)$ over all $r\ge1$, all factors $Z$ of $W_j[r]$ with $\abs Z\le d_j$, and all $x$.

The first lemma says that the pairs at a point are records of the period just before it or just after it.

\begin{lemma}
\label{lem:pairs-records}
Let $p$ be a point $(x,t)$ of $W\coloneqq W_j[r]=w_1\cdots w_m$, and let
\begin{align*}
Z^-\coloneqq w_{t-d_j}\cdots w_{t-1},\qquad Z^+\coloneqq w_{t+1}\cdots w_{t+d_j},
\end{align*}
cut off at the ends of $W$, with the positions of $W$. The positions $s$ of the earlier points of the empty increasing pairs with the later point $p$, and of the later points of those with the earlier point $p$, satisfy
\begin{align}
\bigl\{s:(w_s,s)\in F_L(p)\bigr\}\subseteq R^-(Z^-,x),\qquad\bigl\{s:p\in F_L\bigl((w_s,s)\bigr)\bigr\}\subseteq R^+(Z^+,x).
\label{eq:pairs-records}
\end{align}
In particular $A(W)\le H_j$.
\end{lemma}

\begin{proof}
Let $(y,s)\in F_L(p)$, so $s<t$, $y<x$, and the rectangle of $(y,s)$ and $p$ contains no other search:
\begin{align*}
w_{s'}\notin[y,x]\qquad(s<s'<t).
\end{align*}
This has two consequences. First, $y$ is not searched at the $t-s-1$ positions between $s$ and $t$, while every $d_j$ consecutive positions of $W$ contain every key by \eqref{eq:level-basic}. So $t-s-1<d_j$, that is,
\begin{align*}
s\ge t-d_j,
\end{align*}
and $s$ is a position of $Z^-$. Second, no key in $(y,x)$ is searched after $s$ in $Z^-$, which is the condition $s\in R^-(Z^-,x)$ of \eqref{eq:records}. This proves the first inclusion of \eqref{eq:pairs-records}.

For the second inclusion, let $p\in F_L((y,s))$, so $s>t$, $y>x$, and $w_{s'}\notin[x,y]$ for $t<s'<s$. In the same way $s\le t+d_j$, and no key in $(x,y)$ is searched before $s$ in $Z^+$, so $s\in R^+(Z^+,x)$.

The factors $Z^-$ and $Z^+$ of $W$ have length at most $d_j$, so both sets on the right of \eqref{eq:pairs-records} have at most $H_j$ elements, and $A(W)\le H_j$.
\end{proof}

The rest of this subsection bounds $H_j$ by induction on $j$. Fix $j\ge1$ and $r\ge1$, and let $W\coloneqq W_j[r]=w_1\cdots w_m$, with the strip word $U=W_{j-1}[qr]=P_{j-1}^{qr}$ and the in-strip words $V'=P_{j-1}^r\ P_{j-1}[0]$ of \cref{def:level}. Every key occurs $2^{j-1}$ times in $P_{j-1}$, and $d_j=2qd_{j-1}$ (\cref{lem:level-basic}). For a strip $a$ and a set $I$ of consecutive visits of $W$, let $N_a(I)$ be the number of visits in $I$ to $a$. Since $U$ repeats the period $P_{j-1}$ of length $d_{j-1}$, every $d_{j-1}$ consecutive visits form a cyclic shift of $P_{j-1}$ and thus visit $a$ exactly $2^{j-1}\ge1$ times; consequently every $qd_{j-1}$ consecutive visits visit $a$ exactly $q2^{j-1}=d_{j-1}$ times, and every shorter set of consecutive visits lies inside one of length $qd_{j-1}$ (as $\abs U=qd_{j-1}r\ge qd_{j-1}$). So
\begin{align}
N_a(I)\ge1\quad\text{if }\abs I=d_{j-1},\qquad N_a(I)\le d_{j-1}\quad\text{if }\abs I\le qd_{j-1}.
\label{eq:visit-counts}
\end{align}
The $i$-th visit to a strip $a$ has the first point $(a,V'[i-1])$ and the second point $(a,V'[i])$ (\cref{def:shuffle}), and $V'$ is a prefix of the level word $W_{j-1}[r+1]=P_{j-1}^{r+1}$. Fix also a factor $Z$ of $W$ with $\abs Z\le d_j$, whose positions we take to be those of $W$, and a key $x=(c,h)$ with strip $c$, or $x=\pm\infty$, in which case we put $c\coloneqq x$. For a set $S$ of strips we write $R^-_S$ and $R^+_S$ for the sets of the lower records below $x$ and of the upper records above $x$ of $Z$ that lie in strips of $S$, with $R^\pm_a\coloneqq R^\pm_{\{a\}}$, and $R^-_{<b}$ and $R^+_{>b}$ for $S$ the set of the strips below and above a strip $b$.

The first lemma counts the records in one strip. Its proof restricts $Z$ to the points of that strip, whose labels run through a factor of a level-$(j-1)$ word.

\begin{lemma}
\label{lem:records-strip}
In the setting above, for every strip $a$,
\begin{align*}
\abs{R^-_a}\le2H_{j-1},\qquad\abs{R^+_a}\le2H_{j-1}.
\end{align*}
\end{lemma}

\begin{proof}
We bound $R^-_a$; for $R^+_a$ exchange below and above, $<$ and $>$, and $+\infty$ and $-\infty$ throughout. If $a>c$, every key of the strip $a$ is above $x$, since keys are ordered by strip first, so $R^-_a$ is empty. Otherwise a point $(a,\ell)$ lies below $x$ exactly when $\ell<h_a$, where
\begin{align*}
h_a\coloneqq\begin{cases}h&(a=c),\\+\infty&(a<c).\end{cases}
\end{align*}

We first count the records at first points. Let $I$ be the set of the visits whose first points lie in $Z$. Let $T_a$ be the set of the positions of $Z$ at the first points of the visits to $a$ in $I$. Let $L$ be the word of the labels at the positions of $T_a$, in order of time, and $\iota$ the one-to-one map from $T_a$ to the positions of $L$. The positions of $Z$ alternate between first and second points, so $\abs I\le\ceil{\abs Z/2}\le qd_{j-1}$, and by \eqref{eq:visit-counts}
\begin{align*}
\abs L=N_a(I)\le d_{j-1}.
\end{align*}
The labels of the first points of consecutive visits to $a$ are consecutive letters of $V'$, so $L$ is a factor of length at most $d_{j-1}$ of $W_{j-1}[r+1]$, and by the definition of $H_{j-1}$
\begin{align*}
\abs{R^-(L,h_a)}\le H_{j-1}.
\end{align*}
A record of $Z$ at a first point of $a$ is a record of $L$:
\begin{align*}
\iota\bigl(R^-_a\cap T_a\bigr)\subseteq R^-(L,h_a).
\end{align*}
Indeed, a later letter of $L$ strictly between the label of a record $s$ and $h_a$ would be the label of a later point of $Z$ with a key strictly between $w_s$ and $x$. So at most $H_{j-1}$ records of $a$ are first points.

The second points are counted in the same way: the second points in $Z$ also belong to at most $qd_{j-1}$ consecutive visits, and the word $L'$ of their labels in the strip $a$ is a factor of length at most $d_{j-1}$ of $W_{j-1}[r+1]$. Hence
\begin{align*}
\abs{R^-_a}\le\abs{R^-(L,h_a)}+\abs{R^-(L',h_a)}\le2H_{j-1}.
\end{align*}
\end{proof}

The second lemma counts the records in the strips beyond the nearest strip that $Z$ visits, through the records of the strip word $U$. Let $b^-$ be the largest strip below $c$ that contains a point of $Z$, and $b^+$ the smallest strip above $c$ that contains a point of $Z$, if they exist.

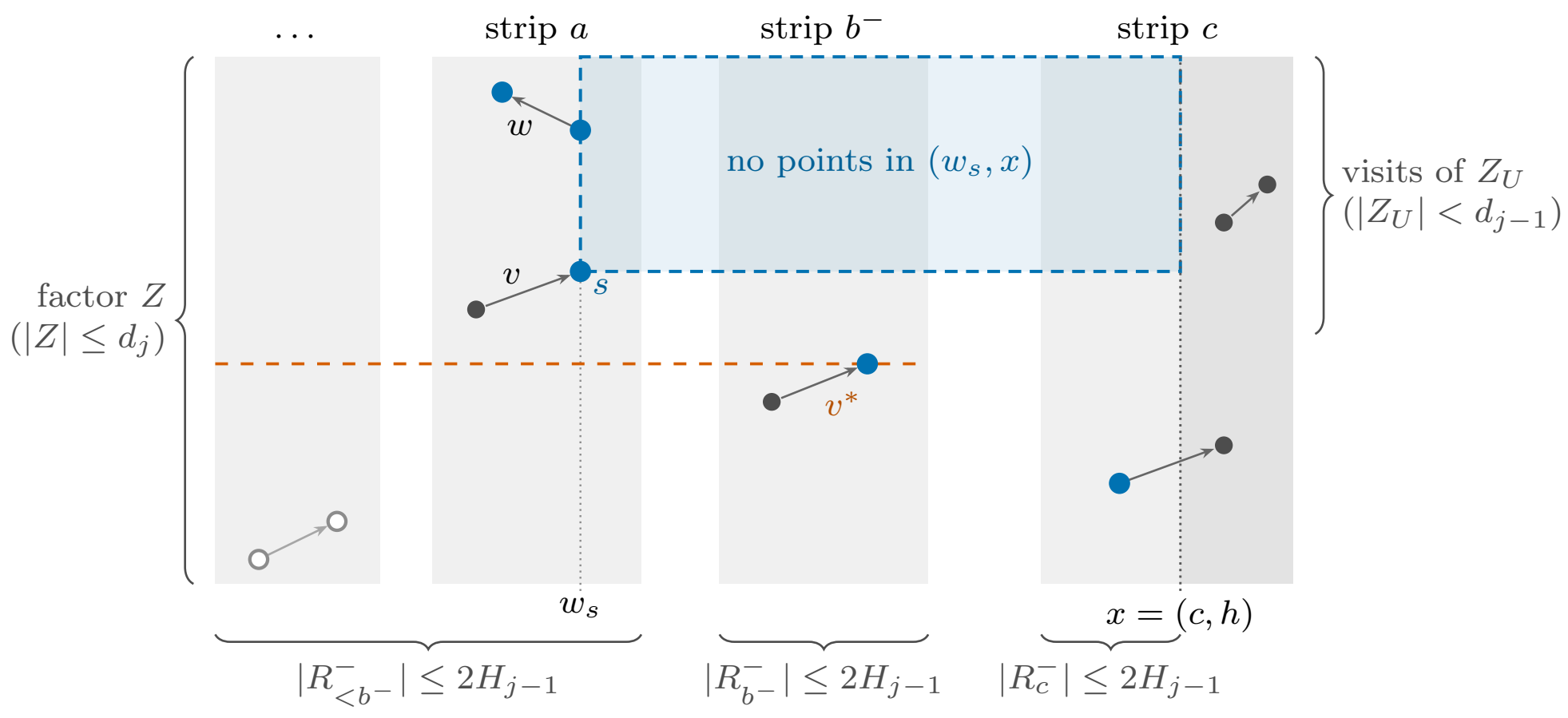
\begin{figure}[t]
\centering
\input{figures/records}
\caption{Bounding the lower records $R^-(Z,x)$ of a factor $Z$ of $W_j[r]$ below $x=(c,h)$ (\cref{lem:records-strip,lem:records-far,lem:records-recursion}). Here $b^-$ is the largest strip below $c$ visited by $Z$ (the strip $b$ immediately below $c$ when $\abs Z=d_j$), and $v^\ast$ is its last visit in $Z$. Earlier points in strips below $b^-$ (open circles) are followed by $v^\ast$, whose keys lie in $(w_s,x)$, so every lower record in strips below $b^-$ comes after $v^\ast$, in the factor $Z_U$ of fewer than $d_{j-1}$ visits. For such a record $s$ in visit $v$ to strip $a<b^-$, later visits in $Z_U$ avoid $(a,b^-)$, although a later visit $w$ may return to $a$ with keys at most $w_s$. The blue dots mark the lower records below $x$, at most $2H_{j-1}$ in each of the three parts.}
\label{fig:records}
\end{figure}

\begin{lemma}
\label{lem:records-far}
In the setting above, if $b^-$ exists, the lower records of $Z$ below $x$ in the strips below $b^-$ satisfy $\abs{R^-_{<b^-}}\le2H_{j-1}$. If $b^+$ exists, the upper records of $Z$ above $x$ in the strips above $b^+$ satisfy $\abs{R^+_{>b^+}}\le2H_{j-1}$.
\end{lemma}

\begin{proof}
We bound $R^-_{<b^-}$; the bound on $R^+_{>b^+}$ is the same with time and the order of the keys reversed, with the first visit to $b^+$ in place of the last visit to $b^-$. Let $v^\ast$ be the last visit to $b^-$ that meets $Z$, and let $Z_U$ be the factor of $U$ that consists of the visits after $v^\ast$ that meet $Z$. Let $s\in R^-_{<b^-}$ lie in the visit $v$ to a strip $a<b^-$. We show
\begin{align}
v^\ast<v,\qquad\abs{Z_U}<d_{j-1},\qquad v\in R^-(Z_U,b^-),
\label{eq:records-far}
\end{align}
where in the last relation the strip $b^-$, a key of $U$, is the threshold.

For the first relation, every key of the strip $b^-$ lies strictly between $w_s$ and $x$:
\begin{align*}
w_s<\bigl(b^-,\ell\bigr)<x\qquad\text{for every label }\ell.
\end{align*}
So no point of $b^-$ follows $s$ in $Z$. The visit $v^\ast$ has a point in $Z$, so it does not come after $v$, and it differs from $v$.

For the second relation, $Z_U$ contains no visit to $b^-$, while every $d_{j-1}$ consecutive visits include one by the first relation of \eqref{eq:visit-counts}.

For the third relation, every visit $w$ in $Z_U$ meets $Z$. If $w$ comes after $v$, its point in $Z$ comes after $s$, so its key is not in $(w_s,x)$, and its strip satisfies
\begin{align*}
u_w\notin(a,b^-).
\end{align*}
This is the condition for $v\in R^-(Z_U,b^-)$. A later visit to the strip $a$ itself is allowed here, by the ties in \eqref{eq:records}, although it may disqualify $s$ in $Z$.

By \eqref{eq:records-far}, $Z_U$ is a factor of length at most $d_{j-1}$ of the level word $U=W_{j-1}[qr]$, and the map that sends $s$ to its visit $v$ maps $R^-_{<b^-}$ into $R^-(Z_U,b^-)$. Each visit has two points, so
\begin{align*}
\abs{R^-_{<b^-}}\le2\abs{R^-(Z_U,b^-)}\le2H_{j-1}.
\end{align*}
\end{proof}

\begin{lemma}
\label{lem:records-recursion}
The numbers $H_j$ satisfy
\begin{align}
H_0\le2,\qquad H_j\le6H_{j-1}\quad(j\ge1),\qquad A\bigl(W_j[r]\bigr)\le H_j\le2\cdot6^j.
\label{eq:level-records}
\end{align}
\end{lemma}

\begin{proof}
For $j=0$, where $d_0=g$, a factor of length at most $g$ of $W_0[r]$ consists of at most two increasing runs of distinct keys. In each run, every key below $x$ except the last is followed by a larger key below $x$, and every key above $x$ except the first is preceded by a smaller key above $x$. So $\abs{R^\pm(Z,x)}\le2$, and $H_0\le2$.

Let $j\ge1$, with $W$, $Z$ and $x$ as above. A lower record lies in a strip at most $c$, and no point of $Z$ lies in a strip strictly between $b^-$ and $c$. So (\cref{fig:records})
\begin{align}
R^-(Z,x)\subseteq R^-_{<b^-}\cup R^-_{b^-}\cup R^-_c,
\label{eq:records-split}
\end{align}
where the first two sets are empty if $b^-$ does not exist and the last is empty if $x$ is $+\infty$. \Cref{lem:records-far} bounds the first of the three sets in \eqref{eq:records-split}, and \cref{lem:records-strip} the other two:
\begin{align*}
\abs{R^-(Z,x)}\le2H_{j-1}+2H_{j-1}+2H_{j-1}=6H_{j-1}.
\end{align*}
In the same way, an upper record lies in a strip at least $c$, and $R^+(Z,x)\subseteq R^+_c\cup R^+_{b^+}\cup R^+_{>b^+}$, so $\abs{R^+(Z,x)}\le6H_{j-1}$. Hence $H_j\le6H_{j-1}$, and with $H_0\le2$ this gives $H_j\le2\cdot6^j$. The bound on $A(W_j[r])$ is \cref{lem:pairs-records}.
\end{proof}

\begin{proposition}
\label{prop:lambda-level}
For all $j\ge0$ and $r\ge1$,
\begin{align*}
\lambda\bigl(W_j[r]\bigr)\le1+\frac2g\sum_{i=1}^j6^i<1+\frac{12\cdot6^j}{5g}.
\end{align*}
\end{proposition}

\begin{proof}
In a cyclic scan, the left funnel of a search of $i>0$ consists of the previous search, of $i-1$, and the left funnel of a search of $0$ is empty, so $\lambda(W_0[r])\le1$. Let $j\ge1$ and $W\coloneqq W_j[r]$, with $U$, $V$ and $q$ as in \cref{def:level}. By \cref{prop:pairs},
\begin{align*}
\Lambda(W)&=\Lambda(U)+q\Lambda(V')+\err(W)\\*
&\le\Lambda(U)+q\Lambda(V)+2\cdot2\cdot6^j\cdot\frac{\abs W}{2g}.
\end{align*}
The inequality uses $\Lambda(V')=\Lambda(V)$ and the number $\abs W/(2g)$ of descending visits from \eqref{eq:level-lengths}, and \eqref{eq:err-degree} with $A(W)\le2\cdot6^j$ from \eqref{eq:level-records}. Dividing by $\abs W=2\abs U=2q\abs V$ gives
\begin{align*}
\lambda(W)\le\frac{\lambda(U)+\lambda(V)}2+\frac{2\cdot6^j}g.
\end{align*}
The words $U$ and $V$ are level-$(j-1)$ words, so by induction both are at most $1+(2/g)\sum_{i=1}^{j-1}6^i$, which proves the first bound. The second is $\sum_{i=1}^j6^i<6^{j+1}/5$.
\end{proof}

\subsection{Independent rectangles of level words}
\label{sec:level-rect}

The lower bound needs no error term. By \cref{lem:compose}, each level keeps the pairs of its pieces, adds one empty decreasing pair for every visit that is an upper point of the set of the strip word and does not descend, and doubles the number of upper points. So the induction carries two numbers per search: the size of the set and the number of its upper points.

\begin{proposition}
\label{prop:rect-level}
For all $j\ge0$ and $r\ge1$, the word $W\coloneqq W_j[r]$ has an independent set $F$ of empty decreasing pairs with
\begin{align}
\frac{\abs{\tops(F)}}{\abs W}\ge\Bigl(1-\frac1g\Bigr)\Bigl(1-\frac1r\Bigr),\qquad\frac{\abs F}{\abs W}\ge\Bigl(1-\frac1g\Bigr)\Bigl(1-\frac1r\Bigr)+\frac j2\Bigl(1-\frac3g+\frac1{g^2}\Bigr).
\label{eq:rect-level}
\end{align}
\end{proposition}

\begin{proof}
Let $W\coloneqq w_1\cdots w_{gr}$ be the cyclic scan $W_0[r]$. For a position $t>g$ with $w_t<g-1$, let $\ell(t)$ be the last position before $t$ with the key $g-1$, which lies in the previous round, and let
\begin{align*}
F\coloneqq\{(\ell(t),t):g<t\le gr,\ w_t<g-1\}.
\end{align*}
Between $\ell(t)$ and $t$ the word searches the keys $0,\dots,w_t-1$, so these are empty decreasing pairs by \eqref{eq:edp}. All their lower points have the key $g-1$, and \eqref{eq:staggered} needs two lower points with different keys, so $F$ is independent by \cref{lem:staggered}. Its pairs have distinct upper points, the positions of the keys below $g-1$ from the second round on, so
\begin{align*}
\abs{F}=\abs{\tops(F)}=(r-1)(g-1)=\Bigl(1-\frac1g\Bigr)\Bigl(1-\frac1r\Bigr)\abs W.
\end{align*}
This is \eqref{eq:rect-level} for $j=0$.

Let $j\ge1$, with $U=W_{j-1}[qr]$, $V=W_{j-1}[r]$ and $q=g^{2^{j-1}}$ as in \cref{def:level}, and let $F_U$ and $F_V$ be the sets that the induction hypothesis gives for $U$ and $V$. The conditions \eqref{eq:edp} and \eqref{eq:staggered} involve only the letters up to the upper points, so $F_V$, with the positions of $V$ counted from $0$, is also an independent set of empty decreasing pairs of the in-strip word $V'$, of which $V$ is a prefix. We apply \cref{lem:compose,lem:compose-indep} with $F_\kappa\coloneqq F_V$ for each of the $q$ strips. With the $\abs W/(2g)$ descending visits of \eqref{eq:level-lengths}, the bounds \eqref{eq:compose-bounds} give a set $F$ with
\begin{align*}
\abs{F}&\ge\abs{F_U}+q\abs{F_V}+\abs{\tops(F_U)}-\frac{\abs W}{2g},\\*
\abs{\tops(F)}&\ge2\abs{\tops(F_U)}.
\end{align*}
Dividing by $\abs W$, which is $2\abs U$ and $2q\abs V$,
\begin{align}
\frac{\abs{\tops(F)}}{\abs W}&\ge\frac{\abs{\tops(F_U)}}{\abs U},\notag\\*
\frac{\abs F}{\abs W}&\ge\frac12\Bigl(\frac{\abs{F_U}}{\abs U}+\frac{\abs{F_V}}{\abs V}\Bigr)+\frac12\Bigl(\frac{\abs{\tops(F_U)}}{\abs U}-\frac1g\Bigr).
\label{eq:rect-recursion}
\end{align}
The induction hypothesis for $U$, whose rounds number $qr\ge r$, gives the first bound of \eqref{eq:rect-level}. Since $qr\ge q\ge g$, it also gives
\begin{align*}
\frac{\abs{\tops(F_U)}}{\abs U}-\frac1g&\ge\Bigl(1-\frac1g\Bigr)^2-\frac1g\\*
&=1-\frac3g+\frac1{g^2}.
\end{align*}
With $\abs{F_U}/\abs U$ and $\abs{F_V}/\abs V$ bounded by the induction hypothesis for $j-1$ and $r'\ge r$ rounds, the second line of \eqref{eq:rect-recursion} gives the second bound of \eqref{eq:rect-level}.
\end{proof}

The gain of about $1/2$ per level comes from the pairs of $F^+$ in \cref{lem:compose}, which use a pair of the strip word a second time. The weights $1/2$ and $1/2$ in \eqref{eq:rect-recursion} and in the recursion for $\lambda$ come from \eqref{eq:lengths}, that is, from the overlap of consecutive visits. With weights that sum to less than one, the second line of \eqref{eq:rect-recursion} would only give a bounded number of pairs per search, and with weights that sum to more than one, the recursion for $\lambda$ would no longer keep $\lambda$ bounded.

\subsection{The separating permutations}
\label{sec:level-main}

The permutations of \cref{thm:main-informal} are the splits of the words $W_k[4]$, with a base $g$ large enough for both propositions.

\begin{theorem}[Formal version of \cref{thm:main-informal}]
\label{thm:main-formal}
Let $k\ge1$, $g\coloneqq3\cdot6^k$, and let $\pi_k$ be the split of the level-$k$ word $W_k[4]$ with this base. Then $\pi_k$ is a permutation of $m\coloneqq2^{k+2}g^{2^k}$ keys with $\log_2\log_2m\le2k+2$. It has an independent set $F$ of empty decreasing pairs with $\abs F\ge(k/2)m$, and
\begin{align*}
\Lambda(\pi_k)\le3m,\qquad\Funnel(\pi_k)\le7m,\qquad\OPT(\pi_k)\ge\Bigl(1+\frac k2\Bigr)m.
\end{align*}
In particular $\OPT(\pi_k)\ge(\log_2\log_2m/28)\cdot\Funnel(\pi_k)$.
\end{theorem}

\begin{proof}
Let $W\coloneqq W_k[4]$. By \eqref{eq:level-basic} its length is $\abs{W}=2^{k+2}g^{2^k}=m$. The split of a word gives each search its own key, so $\pi_k$ is a permutation of $m$ keys. By \cref{prop:lambda-level}, the word $W$ has fewer than two empty increasing pairs per search:
\begin{align*}
\lambda(W)<1+\frac{12\cdot6^k}{5g}=1+\frac45.
\end{align*}
By \cref{prop:rect-level} with $r=4$, the word $W$ has an independent set $F$ of empty decreasing pairs with
\begin{align*}
\frac{\abs F}m&\ge\frac34\Bigl(1-\frac1g\Bigr)+\frac k2\Bigl(1-\frac3g+\frac1{g^2}\Bigr)\\*
&\ge\frac k2+\frac34-\frac{3(2k+1)}{4g}\\*
&\ge\frac k2.
\end{align*}
The second inequality drops $k/(2g^2)$ and collects the terms in $1/g$, and the last uses $g\ge2k+1$. By \cref{lem:split}, $F$ remains an independent set of empty decreasing pairs in $\pi_k$. If $W$ has $n$ distinct keys, then
\begin{align*}
\Lambda(\pi_k)&=\Lambda(W)+m-n\le3m,\\*
\Funnel(\pi_k)&\le m+2\Lambda(\pi_k)\le7m,\\*
\OPT(\pi_k)&\ge m+\abs F\ge\Bigl(1+\frac k2\Bigr)m.
\end{align*}
The identity is \cref{lem:split}, followed by $\Lambda(W)=\lambda(W)m\le2m$. The bound on $\Funnel(\pi_k)$ is \cref{lem:funnel-lambda}, and the bound on $\OPT(\pi_k)$ is \cref{lem:dhikp} for the permutation $\pi_k$ and the set $F$. For the size,
\begin{align*}
\log_2m=k+2+2^k\log_2g&\le k+2+2^k(3k+2)\\*
&\le2^k(3k+4)\\*
&\le2^{2k+2}.
\end{align*}
The first inequality uses $\log_2g=\log_23+k\log_26\le3k+2$, the second $k+2\le2^{k+1}$, and the last $3k+4\le2^{k+2}$. Hence $\log_2\log_2m\le2k+2$, and
\begin{align*}
\OPT(\pi_k)&\ge\frac{k+2}2m\\*
&=\frac{2k+4}{28}\cdot7m\\*
&\ge\frac{\log_2\log_2m}{28}\cdot\Funnel(\pi_k).
\end{align*}
\end{proof}

The second inequality for the size shows more: $\log_2\log_2m\le k+\log_2(3k+4)$, so $k=\log_2\log_2m-O(\log\log\log m)$.

%% file: figures/records.tex
\begin{tikzpicture}[x=0.72cm,y=0.45cm]
\fill[black!6] (0.0,0.55) rectangle (1.9,10.25);
\fill[black!6] (2.5,0.55) rectangle (4.9,10.25);
\fill[black!6] (5.8,0.55) rectangle (8.2,10.25);
\fill[black!6] (9.5,0.55) rectangle (12.4,10.25);
\fill[black!11] (11.1,0.55) rectangle (12.4,10.25);

\node[figtxt,anchor=south] at (0.95,10.38) {$\cdots$};
\node[figtxt,anchor=south] at (3.70,10.38) {strip $a$};
\node[figtxt,anchor=south] at (7.00,10.38) {strip $b^-$};
\node[figtxt,anchor=south] at (10.95,10.38) {strip $c$};

\fill[DPBlue,fill opacity=0.09] (4.2,6.3) rectangle (11.1,10.25);
\draw[DPBlue,line width=0.75pt,densely dashed] (4.2,6.3) rectangle (11.1,10.25);
\node[figtxt,text=DPBlue!85!black] at (7.65,8.3) {no points in $(w_s,x)$};

\draw[black!45,line width=0.45pt,densely dotted] (4.2,0.42) -- (4.2,6.3);
\node[figtxt,anchor=north] at (4.2,0.38) {$w_s$};
\draw[black!65,line width=0.55pt,densely dotted] (11.1,0.42) -- (11.1,10.25);
\node[figtxt,anchor=north] at (11.1,0.38) {$x=(c,h)$};

\draw[DPOrange,line width=0.7pt,dashed] (0.0,4.6) -- (8.2,4.6);

\draw[varrow,draw=black!35] (0.5,1.0) -- (1.4,1.7);
\node[cdot,draw=black!45,minimum size=4.2pt] at (0.5,1.0) {};
\node[cdot,draw=black!45,minimum size=4.2pt] at (1.4,1.7) {};

\draw[varrow] (10.4,2.4) -- (11.6,3.1);
\node[tdot,fill=DPBlue] at (10.4,2.4) {};
\node[sdot] at (11.6,3.1) {};

\draw[varrow] (6.4,3.9) -- (7.5,4.6);
\node[sdot] at (6.4,3.9) {};
\node[tdot,fill=DPBlue] at (7.5,4.6) {};
\node[figtxt,anchor=north west,text=DPOrange!85!black] at (6.95,4.18) {$v^\ast$};

\draw[varrow] (3.0,5.6) -- (4.2,6.3);
\node[sdot] at (3.0,5.6) {};
\node[tdot,fill=DPBlue] at (4.2,6.3) {};
\node[figtxt,anchor=south east] at (3.58,5.98) {$v$};
\node[figtxt,anchor=north west,text=DPBlue!85!black] at (4.28,6.25) {$s$};

\draw[varrow] (11.6,7.2) -- (12.1,7.9);
\node[sdot] at (11.6,7.2) {};
\node[sdot] at (12.1,7.9) {};

\draw[varrow] (4.2,8.9) -- (3.3,9.6);
\node[tdot,fill=DPBlue] at (4.2,8.9) {};
\node[tdot,fill=DPBlue] at (3.3,9.6) {};
\node[figtxt,anchor=north east] at (3.72,9.18) {$w$};

\draw[decorate,decoration={brace,amplitude=4pt},black!70,line width=0.5pt] (-0.25,0.55) -- (-0.25,10.25)
  node[midway,left=5pt,figtxt,align=right] {factor $Z$\\($\abs Z\le d_j$)};

\draw[decorate,decoration={brace,amplitude=4pt},black!70,line width=0.5pt] (12.65,10.25) -- (12.65,5.15)
  node[midway,right=5pt,figtxt,align=left] {visits of $Z_U$\\($\abs{Z_U}<d_{j-1}$)};

\draw[decorate,decoration={brace,mirror,amplitude=3.5pt},black!70,line width=0.5pt] (0.0,-0.38) -- (4.9,-0.38)
  node[midway,below=4pt,figtxt] {$\abs{R^-_{<b^-}}\le 2H_{j-1}$};
\draw[decorate,decoration={brace,mirror,amplitude=3.5pt},black!70,line width=0.5pt] (5.8,-0.38) -- (8.2,-0.38)
  node[midway,below=4pt,figtxt] {$\abs{R^-_{b^-}}\le 2H_{j-1}$};
\draw[decorate,decoration={brace,mirror,amplitude=3.5pt},black!70,line width=0.5pt] (9.5,-0.38) -- (11.1,-0.38)
  node[midway,below=4pt,figtxt] {$\abs{R^-_c}\le 2H_{j-1}$};
\end{tikzpicture}

%% file: sections/07_consequences.tex
\section{Limits and consequences of the separation}
\label{sec:consequences}

This section proves the results of \cref{sec:intro} that surround the main theorem: the gap of \cref{thm:main-formal} is the largest possible (\cref{sec:gap}), the independent rectangle bound and the two signed counts on $\pi_k$ (\cref{sec:irb}), and what rotating the plane gives for the left-signed count (\cref{sec:rotation}).

\subsection{The gap is the largest possible}
\label{sec:gap}

The separation of \cref{thm:main-formal} is the largest possible up to a constant factor. Lecomte and Weinstein observe that Tango trees give such a bound in terms of the alternation bound \cite[Section~1]{lw20}; we combine it with their Theorem~1.

\begin{proposition}[Formal version of \cref{prop:tango-informal}]
\label{prop:tango-formal}
For every permutation $X$ of $m\ge2$ keys, $\OPT(X)\le O\bigl((\Funnel(X)+m)(1+\log\log m)\bigr)$.
\end{proposition}

\begin{proof}
Tango trees serve a sequence of $m$ searches on $n$ keys at cost $O((\IB(X)+n+m)(1+\log\log n))$ \cite[proof of Theorem~3.3]{dhip07}, where $n=m$ for a permutation. Here the interleave bound $\IB(X)$ is defined from a balanced binary search tree $H$ on the keys \cite[Section~2]{dhip07}: for each node $z$ of $H$, list the searches in the subtree of $z$ in order of time, label a search left if its key is $z$ or lies in the left subtree of $z$ and right otherwise, and count the changes between left and right; $\IB(X)$ is the sum of these counts over all nodes $z$. The alternation bound $\Alt_T(X)$ of \cite[Definition~10]{lw20} is defined similarly on a binary tree $T$ whose leaves are the keys: at each internal node, a search below it is left or right according to the subtree containing its key, and the node contributes the number of maximal blocks of equal labels, which is at least the number of changes.

To compare the two bounds, turn $H$ into a leaf-labelled tree $T$ by placing each node $z$ as a leaf between its two subtrees. Writing $\langle A,B\rangle$ for the binary tree with left subtree $A$ and right subtree $B$, set
\begin{align*}
T_z\coloneqq\begin{cases}\langle A_z,T_{z^+}\rangle&(z\text{ has a right child }z^+),\\A_z&(\text{otherwise}),\end{cases}\qquad A_z\coloneqq\begin{cases}\langle T_{z^-},z\rangle&(z\text{ has a left child }z^-),\\z&(\text{otherwise}),\end{cases}
\end{align*}
and $T\coloneqq T_z$ for the root $z$ of $H$. By induction, the leaves of $T_z$ are the keys in the subtree of $z$, each once. If $z$ has a right child, the root of $T_z$ partitions those keys into $z$ together with its left subtree (the leaves of $A_z$) and its right subtree (the leaves of $T_{z^+}$), so it contributes at least the count of $z$ in $\IB(X)$; if $z$ has no right child, that count is $0$. Since every other internal node of $T$ contributes a nonnegative number, $\IB(X)\le\Alt_T(X)$. Hence
\begin{align*}
\OPT(X)&\le O\bigl((\IB(X)+m)(1+\log\log m)\bigr)\\*
&\le O\bigl((\Alt_T(X)+m)(1+\log\log m)\bigr)\\*
&\le O\bigl((\Funnel(X)+m)(1+\log\log m)\bigr).
\end{align*}
The first inequality is the Tango tree bound with $n=m$, the second uses $\IB(X)\le\Alt_T(X)$, and the last applies \cite[Theorem~1]{lw20}, which holds for every leaf-labelled reference tree $T$.
\end{proof}

Every search after the first has the immediately preceding search in its funnel, so $\Funnel(X)\ge m-1\ge m/2$ for $m\ge2$; when $m\ge4$, we also have $\log_2\log_2m\ge1$, so \cref{prop:tango-formal} simplifies to $\OPT(X)\le O(\log\log m)\cdot\Funnel(X)$, which is \cref{prop:tango-informal}.

\subsection{The independent rectangle bound and the signed counts}
\label{sec:irb}

This subsection proves \cref{cor:irb-formal}, the formal version of \cref{cor:irb-informal}. The independent rectangle bound $\IRB(X)$ of a permutation $X$ is the largest size of a set of pairwise independent empty rectangles spanned by two searches, of either orientation \cite[Section~4.1]{dhi+09}; an independent set of empty decreasing pairs (\cref{def:independent}) is such a set.

The signed counts are those of the signed Greedy algorithm \cite[Algorithm~4.3]{dhi+09}, described in \cref{sec:intro}: $\add(X)$ is the number of points that it adds for the rectangles whose lower corner lies to the right, and $\add_L(X)$ the number for the mirror orientation. Demaine et al.\ prove the following two facts for permutations.

\begin{lemma}[{\cite[Lemma~4.2]{dhi+09}}]
\label{lem:greedy-points}
For every permutation $X$ of $m$ keys, the least size of a right-signed satisfied set $S\supseteq\Pts(X)$ is $m+\add(X)$.
\end{lemma}

Together with \cref{lem:dhikp}, which is \cite[Lemma~4.5]{dhi+09} and bounds the same least size from below by $m+\abs F$, this gives, for every independent set $F$ of empty decreasing pairs of a permutation $X$,
\begin{align}
\add(X)\ge\abs F.
\label{eq:add-F}
\end{align}

\begin{lemma}[{\cite[Lemma~4.1]{dhi+09}}]
\label{lem:left-greedy}
For every permutation $X$, there is an independent set of $\add_L(X)$ rectangles, each spanned by two searches of which the lower one lies to the left, and each containing no other search.
\end{lemma}

Each of these rectangles is spanned by an empty increasing pair, and distinct rectangles have distinct pairs, so
\begin{align}
\add_L(X)\le\Lambda(X).
\label{eq:addL-Lambda}
\end{align}

\begin{corollary}[Formal version of \cref{cor:irb-informal}]
\label{cor:irb-formal}
The permutations $\pi_k$ of \cref{thm:main-formal} satisfy
\begin{align*}
\IRB(\pi_k)\ge\frac k2m,\qquad\add(\pi_k)\ge\frac k2m,\qquad\add_L(\pi_k)\le3m.
\end{align*}
\end{corollary}

\begin{proof}
Let $F$ be the independent set of empty decreasing pairs of $\pi_k$ of \cref{thm:main-formal}. Its rectangles are pairwise independent empty rectangles, so $\IRB(\pi_k)\ge\abs F$, and $\add(\pi_k)\ge\abs F$ by \eqref{eq:add-F}. With \eqref{eq:addL-Lambda},
\begin{align*}
\IRB(\pi_k)\ge\abs F\ge\frac k2m,\qquad\add(\pi_k)\ge\abs F\ge\frac k2m,\qquad\add_L(\pi_k)\le\Lambda(\pi_k)\le3m.
\end{align*}
\end{proof}

With $\Funnel(\pi_k)\le7m$ from \cref{thm:main-formal}, \cref{cor:irb-formal} gives $\IRB(\pi_k)\ge(k/14)\Funnel(\pi_k)$ and $\add(\pi_k)-\add_L(\pi_k)\ge(k/2-3)m$. Since $k\ge(\log_2\log_2m)/2-1$ by the same theorem, these are the statements of \cref{cor:irb-informal}.

\subsection{Rotating the plane}
\label{sec:rotation}

Rotating the plane by $90$ degrees turns empty decreasing pairs into empty increasing pairs, and so a lower bound on the right-signed count into one on the left-signed count. As a first step towards showing that the funnel bound matches the independent rectangle bound, Lecomte and Weinstein proved that the left-signed count is at most $O(m)$ plus $m$ times the number of empty occurrences of the pattern $3142$, and hence $O(m)$ when there are $O(1)$ such occurrences \cite[Theorems~3 and~34]{lw20}; here an occurrence consists of four searches whose keys, in order of time, are ordered as $3,1,4,2$ and whose bounding rectangle contains no other search. On $\pi_k$ the left-signed count is $O(m)$ as well (\cref{cor:irb-formal}), and a rotation moves the separation from the right-signed count to the left-signed count. For a permutation $X$ of $1,\dots,m$, let $\rho X$ be the permutation whose point set is the image of $\Pts(X)$ under
\begin{align*}
\rho(x,t)\coloneqq(m+1-t,\ x).
\end{align*}
\begin{corollary}
\label{cor:rotation}
The permutations $\rho\pi_k$ have at most $3.5m$ empty occurrences of the pattern $3142$, and $\add_L(\rho\pi_k)\ge(k/2)m$.
\end{corollary}

\begin{proof}
The number of empty occurrences of $3142$ is invariant under $\rho$ \cite[Lemma~22]{lw20} and at most half the funnel bound \cite[Lemma~24]{lw20}, so it is at most $3.5m$ for $\rho\pi_k$ by \cref{thm:main-formal}.

For the left-signed count, let $X\coloneqq\pi_k$, with its keys renamed $1,\dots,m$ in increasing order, and let $\bar X$ be $\rho X$ with the order of the keys reversed, so that $\add_L(\rho X)=\add(\bar X)$ by the definition of the left-signed count. Reversing the keys maps the point $(m+1-t,x)$ to $(t,x)$, so
\begin{align*}
\Pts(\bar X)=\{(t,x_t):1\le t\le m\},
\end{align*}
the image of $\Pts(X)$ under the transposition $(x,t)\mapsto(t,x)$. The transposition maps rectangles onto rectangles, corners to corners and interiors to interiors. For an empty decreasing pair $(s,t)$ of $X$, the images $(s,x_s)$ and $(t,x_t)$ of its two points satisfy
\begin{align*}
x_t<x_s,\qquad t>s,
\end{align*}
so the image of the upper point is the earlier of the two and has the larger key, and $(x_t,x_s)$ is an empty decreasing pair of $\bar X$ whose rectangle is the image of $B(s,t)$. Hence the set $F$ of \cref{thm:main-formal} is mapped to an independent set $\bar F$ of empty decreasing pairs of $\bar X$ with $\abs{\bar F}=\abs F$. By \eqref{eq:add-F},
\begin{align*}
\add_L(\rho\pi_k)&=\add(\bar X)\\*
&\ge\abs{\bar F}=\abs F\\*
&\ge\frac k2m.
\end{align*}
\end{proof}

Lecomte and Weinstein note that if the funnel bound matched the independent rectangle bound, their bound could be improved to $O(m)$ plus a constant times the number of occurrences \cite[remark after Theorem~34]{lw20}. \Cref{cor:rotation} shows that this improvement fails: on $\rho\pi_k$ the number of occurrences is $O(m)$, while the left-signed count is $\Omega(m\log\log m)$.

\begin{remark}
\label{rem:symmetries}
The funnel bound stays $O(m)$ on every image of $\pi_k$ under the eight symmetries of the square, which map permutations to permutations and do not change $\OPT$ \cite[Section~4.1]{dhi+09}. Mirroring the keys exchanges left and right funnels and leaves every $f(p)$ unchanged. For rotations by $90$ degrees, let $\mathrm{zRects}(P)$ be the number of empty occurrences of the pattern $3142$ in a point set $P$, the z-rectangles of \cite[Definitions~19 and~20]{lw20}. It is invariant under rotation \cite[Lemma~22]{lw20}, and $\Funnel(P)/2-O(m)\le\mathrm{zRects}(P)\le\Funnel(P)/2$ \cite[Lemmas~23 and~24]{lw20}; these lemmas are stated for point sets with distinct coordinates in both axes, which the images of a permutation are. So symmetrizing the funnel bound does not close the gap.
\end{remark}

%% file: sections/11_ai_methodologies.tex
\section*{AI methodology statement}

We used Claude Opus 5.5 and Claude Fable 5.2 (Anthropic), GPT-6.2 Astra (OpenAI) and Gemini 4 Carbon (Google) throughout this work, including the development of the construction and its proofs and the writing of the manuscript. The authors are responsible for all content.

%% file: main.bbl
\newcommand{\etalchar}[1]{$^{#1}$}
\begin{thebibliography}{CGK{\etalchar{+}}16}

\bibitem[CCS23]{ccs23}
Parinya Chalermsook, Julia Chuzhoy, and Thatchaphol Saranurak.
\newblock Pinning down the strong {W}ilber-1 bound for binary search trees.
\newblock {\em Theory of Computing}, 19(8):1--71, 2023.
\newblock Preliminary version in APPROX/RANDOM 2020.

\bibitem[CGK{\etalchar{+}}15]{cgk+15}
Parinya Chalermsook, Mayank Goswami, L{\'a}szl{\'o} Kozma, Kurt Mehlhorn, and
  Thatchaphol Saranurak.
\newblock Pattern-avoiding access in binary search trees.
\newblock In {\em Proceedings of the 56th Annual IEEE Symposium on Foundations
  of Computer Science (FOCS)}, pages 410--423, 2015.
\newblock Numbering of appendices as in arXiv:1507.06953v1.

\bibitem[CGK{\etalchar{+}}16]{cgk+16}
Parinya Chalermsook, Mayank Goswami, L{\'a}szl{\'o} Kozma, Kurt Mehlhorn, and
  Thatchaphol Saranurak.
\newblock The landscape of bounds for binary search trees.
\newblock arXiv:1603.04892, 2016.

\bibitem[CHH{\etalchar{+}}26]{chh+26}
Petr Chmel, Bernhard Haeupler, Richard Hlad{\'\i}k, Michal Kouck{\'y}, Antti
  Roeyskoe, V{\'a}clav Rozho{\v{n}}, Ond{\v{r}}ej Sladk{\'y}, and Robert~E.
  Tarjan.
\newblock Splay trees are almost dynamically optimal.
\newblock arXiv:2607.18498, 2026.

\bibitem[Der09]{der09}
Jonathan~C. Derryberry.
\newblock {\em Adaptive Binary Search Trees}.
\newblock PhD thesis, Carnegie Mellon University, 2009.
\newblock Technical Report CMU-CS-09-180.

\bibitem[DHI{\etalchar{+}}09]{dhi+09}
Erik~D. Demaine, Dion Harmon, John Iacono, Daniel Kane, and Mihai
  P{\u{a}}tra{\c{s}}cu.
\newblock The geometry of binary search trees.
\newblock In {\em Proceedings of the 20th Annual ACM-SIAM Symposium on Discrete
  Algorithms (SODA)}, pages 496--505, 2009.

\bibitem[DHIP07]{dhip07}
Erik~D. Demaine, Dion Harmon, John Iacono, and Mihai P{\u{a}}tra{\c{s}}cu.
\newblock Dynamic optimality---almost.
\newblock {\em SIAM Journal on Computing}, 37(1):240--251, 2007.
\newblock Preliminary version in FOCS 2004.

\bibitem[DSW05]{dsw05}
Jonathan Derryberry, Daniel~Dominic Sleator, and Chengwen~Chris Wang.
\newblock A lower bound framework for binary search trees with rotations.
\newblock Technical Report CMU-CS-05-187, Carnegie Mellon University, 2005.

\bibitem[GG19]{gg19}
Navin Goyal and Manoj Gupta.
\newblock Better analysis of {G}reedy binary search tree on decomposable
  sequences.
\newblock {\em Theoretical Computer Science}, 776:19--42, 2019.

\bibitem[GPSW26]{gpsw26}
Yuhao Guo, Seth Pettie, Daniel Skora, and Chengzhang Wan.
\newblock The {G}reedy binary search tree is non-trivially competitive.
\newblock arXiv:2608.04410, 2026.
\newblock To appear in SODA 2027.

\bibitem[Iac05]{iac05}
John Iacono.
\newblock Key-independent optimality.
\newblock {\em Algorithmica}, 42(1):3--10, 2005.

\bibitem[Iac13]{iac13}
John Iacono.
\newblock In pursuit of the dynamic optimality conjecture.
\newblock In {\em Space-Efficient Data Structures, Streams, and Algorithms:
  Papers in Honor of {J.} {I}an {M}unro on the Occasion of His 66th Birthday},
  volume 8066 of {\em Lecture Notes in Computer Science}, pages 236--250.
  Springer, 2013.

\bibitem[JLWY24]{jlwy24}
Shunhua Jiang, Victor Lecomte, Omri Weinstein, and Sorrachai
  Yingchareonthawornchai.
\newblock Hardness amplification for dynamic binary search trees.
\newblock In {\em Proceedings of the 35th International Symposium on Algorithms
  and Computation (ISAAC)}, volume 322 of {\em LIPIcs}, pages 42:1--42:19,
  2024.

\bibitem[LT19]{lt19}
Caleb~C. Levy and Robert~E. Tarjan.
\newblock A new path from {S}play to dynamic optimality.
\newblock In {\em Proceedings of the 30th Annual ACM-SIAM Symposium on Discrete
  Algorithms (SODA)}, pages 1311--1330, 2019.

\bibitem[Luc88]{luc88}
Joan~M. Lucas.
\newblock Canonical forms for competitive binary search tree algorithms.
\newblock Technical Report DCS-TR-250, Rutgers University, 1988.

\bibitem[LW20]{lw20}
Victor Lecomte and Omri Weinstein.
\newblock Settling the relationship between {W}ilber's bounds for dynamic
  optimality.
\newblock In {\em Proceedings of the 28th Annual European Symposium on
  Algorithms (ESA)}, volume 173 of {\em LIPIcs}, pages 68:1--68:21, 2020.
\newblock Numbering of results as in the full version, arXiv:1912.02858v3.

\bibitem[Mun00]{mun00}
J.~Ian Munro.
\newblock On the competitiveness of linear search.
\newblock In {\em Proceedings of the 8th Annual European Symposium on
  Algorithms (ESA)}, volume 1879 of {\em Lecture Notes in Computer Science},
  pages 338--345, 2000.

\bibitem[ST85]{st85}
Daniel~Dominic Sleator and Robert~Endre Tarjan.
\newblock Self-adjusting binary search trees.
\newblock {\em Journal of the ACM}, 32(3):652--686, 1985.

\bibitem[WDS06]{wds06}
Chengwen~Chris Wang, Jonathan Derryberry, and Daniel~Dominic Sleator.
\newblock {$O(\log\log n)$}-competitive dynamic binary search trees.
\newblock In {\em Proceedings of the 17th Annual ACM-SIAM Symposium on Discrete
  Algorithms (SODA)}, pages 374--383, 2006.

\bibitem[Wil89]{wil89}
Robert Wilber.
\newblock Lower bounds for accessing binary search trees with rotations.
\newblock {\em SIAM Journal on Computing}, 18(1):56--67, 1989.
\newblock Preliminary version in FOCS 1986.

\end{thebibliography}
